\documentclass[structabstract]{aa}
\usepackage{graphicx}
\usepackage{rotating}
\usepackage{txfonts}
\usepackage{natbib}
\usepackage{lscape}
\usepackage{booktabs}
\usepackage{dcolumn}
\usepackage{float}
\usepackage{dblfloatfix}

\usepackage{color}
\usepackage{hyperref}
\hypersetup{
  colorlinks=true,        % false: boxed links; true: colored links
  linkcolor=blue,         % color of internal links
  citecolor=cyan,         % color of links to bibliography
}
\bibpunct{(}{)}{;}{a}{}{,} % to follow the A&A style 

\begin{document}
\renewcommand{\topfraction}{1.99}
\renewcommand{\bottomfraction}{1.99}
\renewcommand{\floatpagefraction}{1.99}
\renewcommand{\dbltopfraction}{1.99}
\renewcommand{\dblfloatpagefraction}{1.99}
\setcounter{totalnumber}{10}
\renewcommand{\textfraction}{1.99}

%---- LR's defs --------------
\newcommand{\cf}{cf.,~}
\newcommand{\ie}{i.e.,~}
\newcommand{\eg}{e.g.,~}
\newcommand{\etal}{\textit{et al.},}
\newcommand{\lr}[1]{\textcolor{red}{LR: #1}}
%---- LR's defs --------------

\newcommand{\cmf}[1]{\textcolor{green}{CF: #1}}
\newcommand{\zy}[1]{\textcolor{green}{ZY: #1}}

\def\simless{\mathbin{\lower 3pt\hbox
{$\rlap{\raise 5pt\hbox{$\char'074$}}\mathchar"7218$}}}   %< or of order
\def\simmore{\mathbin{\lower 3pt\hbox
{$\rlap{\raise 5pt\hbox{$\char'076$}}\mathchar"7218$}}}   %> or of order

\title{Jet-torus connection in radio galaxies}
\subtitle{Relativistic hydrodynamics and synthetic emission}
\author{C. M. Fromm\inst{1,2}, M. Perucho\inst{3,4}, O. Porth\inst{1},
  Z. Younsi\inst{1}, E. Ros\inst{2,3,4}, Y. Mizuno\inst{1},
  J. A. Zensus\inst{2} \and L. Rezzolla\inst{1,5}}
\institute{Institut f\"ur Theoretische Physik, Goethe Universit\"at,
  Max-von-Laue-Str. 1, D-60438 Frankfurt, Germany\
\and Max-Planck-Institut f\"ur Radioastronomie, Auf dem H\"ugel 69,
D-53121 Bonn, Germany\
\and Departament d'Astronomia i Astrof\'\i sica, Universitat de
Val\`encia, Dr. Moliner 50, E-46100 Burjassot, Val\`encia, Spain\
\and Observatori Astron\`omic, Parc Cient\'{\i}fic, Universitat de
Val\`encia, C/ Catedr\`atic Jos\'e Beltr\'an 2, E-46980 Paterna,
Val\`encia, Spain 
\and Frankfurt Institute for Advanced Studies, Ruth-Moufang-Strasse 1,
60438 Frankfurt, Germany 
\email{cfromm@th.physik.uni-frankfurt.de}}
%\date{Received September 15, 1996; accepted March 16, 1997}
% \abstract{}{}{}{}{} 
% 5 {} token are mandatory
\abstract
% context heading (optional) 
    {High-resolution Very-Long-Baseline Interferometry observations of
      active galactic nuclei have revealed asymmetric structures in the
      jets of radio galaxies. These asymmetric structures may be due to
      internal asymmetries in the jet, could be induced by the different
      conditions in the surrounding ambient medium including the
      obscuring torus, or a combination of the two.}
% aims heading (mandatory)
   {In this paper we investigate the influence of the ambient medium
     (including the obscuring torus) on the observed properties of jets
     from radio galaxies.}
% methods heading (mandatory)
   {We performed special-relativistic hydrodynamic (RHD) simulations of
     over-pressured and pressure-matched jets using the
     special-relativistic hydrodynamics code \texttt{Ratpenat}, which is
     based on a second-order accurate finite-volume method and an
     approximate Riemann solver. Using a newly developed emission code
     to compute the electromagnetic emission, we have investigated the
     influence of different ambient medium and torus configurations on
     the jet structure and subsequently computed the non-thermal
     emission produced by the jet and the thermal absorption due to the
     torus. To better compare the emission simulations with observations
     we produced synthetic radio maps, taking into account the properties
     of the observatory.}
% results heading (mandatory)
   {The detailed analysis of our simulations shows that the observed
     asymmetries can be produced by the interaction of the jet with the
     ambient medium and by the absorption properties of the obscuring
     torus.}
% conclusions heading (optional) 
   {} 

\titlerunning{Jet-torus connection in radio galaxies}
\authorrunning{C. M. Fromm et al.}

\maketitle
\section{Introduction}

Active galactic nuclei (AGN) are generally assumed to consist of a
central, compact massive object (\ie a supermassive black hole), an
accretion disk surrounded by a molecular torus, and a jet launched
perpendicular to the black-hole disk system. Within this standard model,
AGN may be classified according to their orientation with respect to
the line of sight. If the AGN is seen at a viewing angle of
$\vartheta=90^\circ$ (\ie edge-on), the innermost regions are obscured by
the torus and more internal features become observable as the viewing
angle decreases \citep{1993ARA&A..31..473A,1995PASP..107..803U}.

In addition to the classification of AGN by their viewing angle,
\citet{1974MNRAS.167P..31F} introduced a division based on the morphology
of the jets themselves: Fanaroff-Riley I (FR I) jets are characterised by
a bright core and FR II jets are dominated by their bright lobes.
 
The large-scale structure of AGN observed at high viewing angles can be
resolved through radio interferometric observations \citep[see,
  \eg][]{2008ApJS..174...74K} and high resolution very-long-baseline
interferometry (VLBI) observations enable us to probe its morphological
evolution on parsec scales \citep[see, \eg][]{2009AJ....137.3718L}. These
jets remain collimated over long distances, up to kilo-parsec scales, at
which they can be modelled using special-relativistic
(magneto-)hydrodynamical [R(M)HD] simulations and the computed emission
and polarisation properties provide insights into the underlying fluid
properties and magnetic field configurations \citep[see,
  \eg][]{2007MNRAS.382..526P, 2002MNRAS.336..328L, 2013MNRAS.430..174H}.

In the infrared band, the electromagnetic emission from these AGN is
assumed to be generated by the re-emission by the obscuring torus of the
radiation initially produced in the accretion disc. Based on
observations, two kinds of torus model were constructed, namely smooth
torus models and clumpy torus models \citep[for a review of torus models
  see, \eg][]{2013arXiv1301.1349H}.

So far both models have been applied successfully to explain
observational results either at kilo-parsec scales (via jet simulations)
or at sub-parsec scales (via torus simulations). However, to explain the
VLBI observations of AGN viewed almost edge-on
$\left(\vartheta\sim90^\circ\right)$, both a jet model and a torus model
should be included. In this paper we combine both approaches, addressing
the question of the impact that the combination of the torus and of the
ambient medium have on the observed properties of the jet at
parsec scales.

Additionally, we aim to make the connection between numerical simulations
of relativistic outflows and their emission with observations in a deeper
way than has been done so far. Jet acceleration, re-collimation, and
other phenomena have been understood with the help of one-dimensional
(1D), 2D, and 3D numerical simulations, including different ingredients
such as magnetic fields and special relativity \cite[see,
  \eg][]{Mizuno2015}. Some emission codes have even been applied
\citep{1997ApJ...482L..33G,Mimica:2009de} to explain the emission. Here
we go one step further and consider additional components in the AGN
picture, such as the obscuring torus, and including time-delays, emission
and absorption in a code which will enable us to compare with real
observations of two-sided sources in future works.

The organisation of the paper is as follows. In Sec.~\ref{setup} we
introduce our numerical setup for the jet and the torus. The results of
the simulations and the emission calculations are presented in
Sects.~\ref{hydro}--\ref{emission}. The discussion of our results is
provided in Sec.~\ref{disc}. Throughout the paper we assume an
ideal-fluid equation of state $p=\rho\epsilon(\hat{\gamma}-1)$, where $p$
is the pressure, $\rho$ the rest-mass density, $\epsilon$ the specific
internal energy, and $\hat{\gamma}$ the adiabatic index
\citep{Rezzolla_book:2013}.

\section{RHD Simulations}
\label{setup}

We performed several 2D axisymmetric simulations of supersonic
special-relativistic hydrodynamical jets using the finite-volume code
\texttt{Ratpenat} \citep[for more details see][and references
  therein]{Perucho:2010ht}.

\subsection{Jet model}
The numerical setup is similar to the one used in
\citet{2016A&A...588A.101F} and for completeness we provide below a short
summary.

The numerical grid consisted of 320 cells in the radial direction and 400
cells in the axial direction. Using a numerical resolution of 4 cells per
jet radius ($R_j$), the grid covered $80\, R_j \times 100\,R_j$. We
assume the $z$ axis to be in the direction of propagation of the jet and
the $x$ axis to represent the radial cylindrical coordinate. The boundary
conditions were reflection at the jet axis, injection at the jet nozzle
and outflow conditions elsewhere. We filled the simulation box with a jet
and the parameters used were: the velocity of the jet $v_{\rm b}$, the
bulk Lorentz factor $\Gamma$, the classical Mach number of the jet $M$,
the rest-mass density of the jet $\rho_{\rm b}$, and the adiabatic index
${\hat \gamma}$. The pressure, $p_{\rm b}$ at the jet nozzle was computed
from the Mach number using an ideal-fluid equation of state. The ratio
$d_{\rm k} = p_{\rm b}/p_{\rm a}$ between the pressure in the jet,
$p_{\rm b}$, and the pressure in the ambient medium, $p_{\rm a}$, led to
a pressure matched (PM) jet (\ie with $d_{\rm k}=1$) or to an
over-pressured (OP) jet (\ie with $d_{\rm k}>1$). Depending on this
ratio, internal structures in the jet, the so-called recollimation
shocks, were formed in the case of an OP jet or the jet appeared feature
less in the case of a PM jet \citep{Mizuno2015}. The properties of the
ambient medium (\eg the gradient in the ambient pressure and ambient
rest-mass density) played a crucial role on the shape of the jet and on
the properties of the created recollimation shocks in the case of OP
jets.  We modelled the decrease in the ambient medium pressure using a
pressure profile presented in \citet{1997ApJ...482L..33G} as
\begin{equation}
p_{\rm a}(z)=\frac{p_{\rm b}}{d_{\rm k}}\left[1+
\left(\frac{z}{z_{\rm c}}\right)^n\right]^{-\frac{m}{n}}\,,
\label{pamb}
\end{equation}
where $z_{\rm c}$ can be considered as the core radius and the exponents
$n$ and $m$ control the steepening of the ambient pressure.  The initial
conditions, given in code units (speed of light $c=1$, jet radius $R_{\rm
  j}$, and ambient medium rest-mass density, $\rho_{\rm a}=1\,{\rm
  g\ cm}^{-3}$) are listed in Table~\ref{jpara} 

\begin{figure}[h!]
\resizebox{\hsize}{!}{\includegraphics{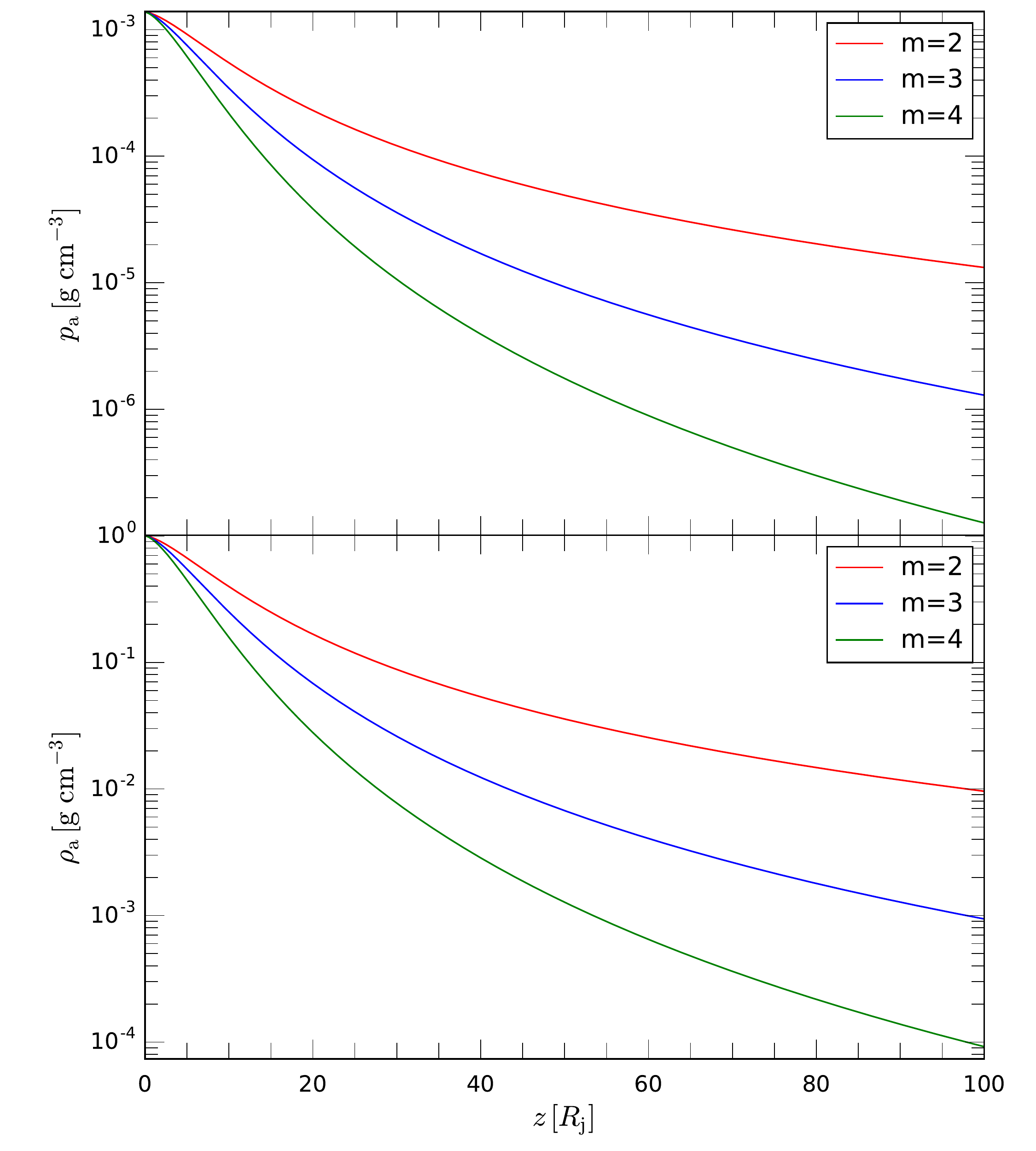}} 
\caption{Distribution of the ambient pressure and the ambient rest-mass
  density in code units for the gradients in the ambient
  medium}
\label{ambient} 
\end{figure}

\begin{table}[h!]
\caption{Initial parameters for the jet.}
\label{jpara}
\centering
\begin{tabular}{c c c c c c c c}
\hline\hline
Model & $R_{\mathrm{b}}$ &	$v_{\mathrm{b}}$ 	&	$d_{\rm k}$	&	$\Gamma$ 	&	$\rho_{\rm b}$		&	$M$	& $\hat\gamma$\\ 
& $[R_{\rm j}]$	&		&		&			& [${\rm g\ cm}^{-3}$] &			&	 \\
\hline
PM	&1		&  0.5	&	1.0		&	1.15	&	0.04&	1.6		&$13/9$\\
OP	&1		&  0.5	&	2.5		&	1.15	&	0.04&	1.6		&$13/9$\\
\hline
\end{tabular}
\end{table}

The distributions of the ambient pressure and of the ambient rest-mass
density for the parameters for the jet models considered here is shown in
Fig.~\ref{ambient}. Throughout the paper we used a single value of
$n=1.5$, while setting $m=2,\,3,\,4$.

\subsection{Torus model}
\label{torusmod}

Once a stationary state for the jet was reached, typically after $5$ grid
longitudinal crossing times, we inserted a steady-state torus. We assumed
that the torus did not influence the dynamics of the jet and therefore
the torus should be regarded as a phenomenological model; we refer the
reader to \citet{2016MNRAS.458.2288S} for a detailed discussion of the
torus modelling.

\begin{figure}[h!]
\begin{center}
\resizebox{0.7\hsize}{!}{\includegraphics{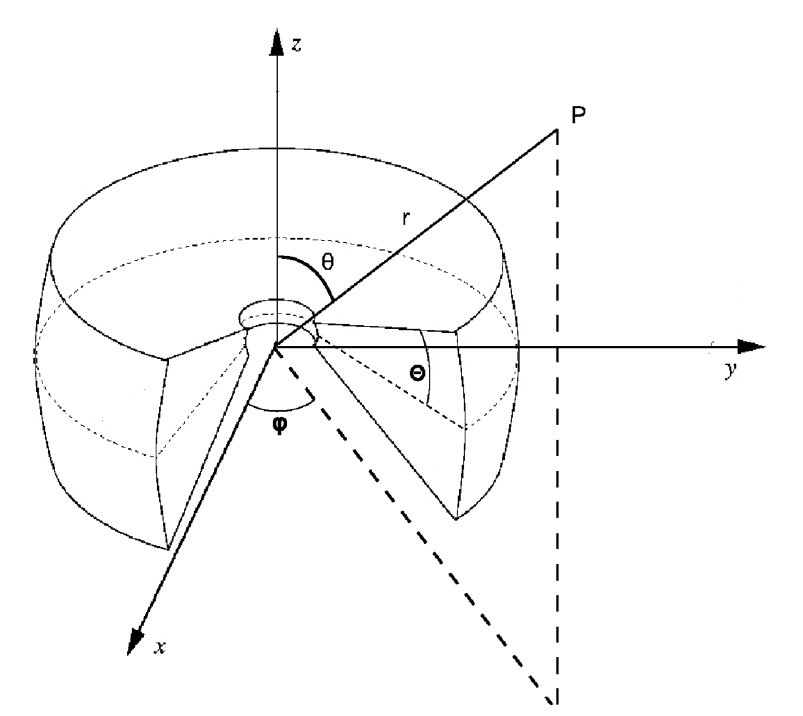}} 
\caption{Geometry of the assumed torus model \citep[adapted
    from][]{2012MNRAS.420.2756S}.}
\label{torussketch} 
\end{center}
\end{figure}

More specifically, our torus model is adapted from
\citet{2012MNRAS.420.2756S,2016MNRAS.458.2288S} and the geometry of the
torus can be seen in Fig. \ref{torussketch}. The geometry of the model
resembles a ``flared disk'' and is characterised by three parameters: the
inner radius of the torus, $R_{\rm in}$, the outer radius of the
torus, $R_{\rm out}$ and the half-opening angle $\Theta$. For the
inner radius of the torus we used
\begin{equation}
\left(\frac{R_{\rm in}}{\mathrm{pc}}\right)\simeq
1.3\left(\frac{L_{\rm AGN}}{10^{46}\mathrm{erg\,s^{-1}}}\right)^{0.5}
\left(\frac{T_{\rm sub}}{1500\mathrm{K}}\right)^{-2.8}\,,
\label{rinner}
\end{equation}
where $L_{\rm AGN}$ is the bolometric luminosity and $T_{\rm sub}$
is the sublimation temperature of the dust grains. We assumed a power law
distribution for the rest-mass density in the torus, given by
\begin{equation}
\rho=\rho \left( R_{\rm in}\right) \left(\frac{r}{R_{\rm in}}
\right)^{-k}e^{-l\left| \cos{\Theta}\right|} \,,
\label{rhotorus}
\end{equation}
where the $\cos\Theta$-dependence in the Eq. (\ref{rhotorus}) leads to a
concentration of rest-mass density towards the equatorial plane. Since we
do not perform a radiative modelling of the torus, its temperature
distribution is simply assumed to follow a certain prescription. In
particular, following \citet{2005A&A...437..861S}, the temperature
decrease moving outwards in the radial direction; furthermore, because of
the direct illumination of the inner torus walls, the temperature should
decrease when going from high to low latitudes. Such behaviour may be
modelled with the following temperature profile
\begin{equation}
T=T_{\rm sub} \left(\frac{r}{R_{\rm in}}\right)^{-k}
{\rm e}^{-l\left|
  \sin{\left(\Theta-\theta\right)}\right|} \,.
\label{temptorus}
\end{equation}

\begin{figure}[h!]
\resizebox{\hsize}{!}{\includegraphics{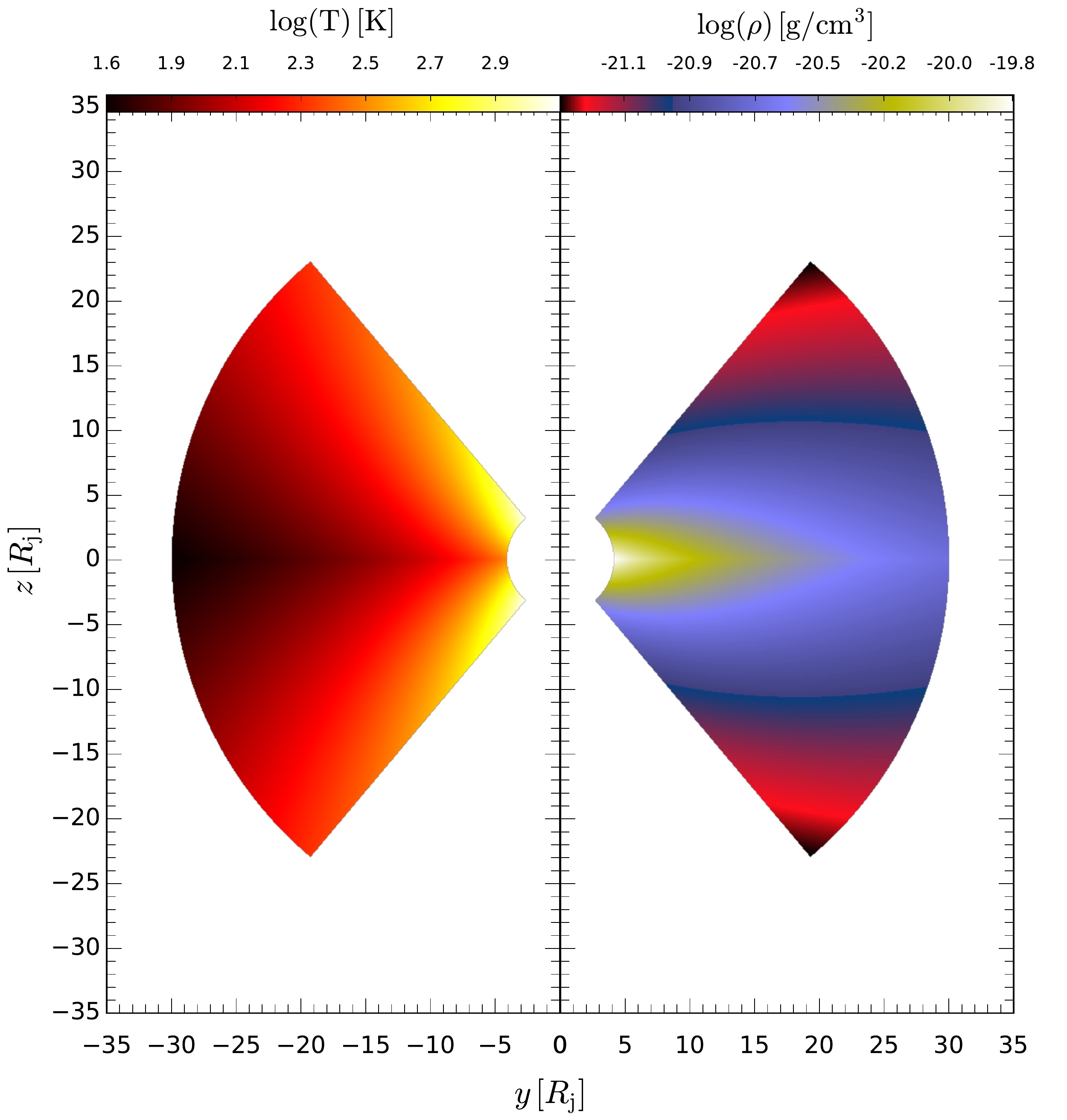}} 
\caption{Left panel: distribution of temperature in the meridional
  plane. Right panel: corresponding rest-mass density distribution. The
  $\Theta$-dependence in the distribution of the temperature and the
  rest-mass density is clearly visible.}
\label{toruspara} 
\end{figure}

Figure~\ref{toruspara} illustrates the distribution of the temperature
and of the rest-mass density in the torus with
$L_{\rm AGN}=10^{43}\,\mathrm{erg\,s^{-1}}$,
$T_{\rm sub}=1500\,\mathrm{K}$, which leads to an inner radius of
$R_{\rm in}=4\,R_{\rm j}$. The outer radius is set to
$R_{\rm out}=30\,R_{\rm j}$ and the angular thickness of the torus
is $\theta=50^\circ$. For the scaling of the rest-mass density we have
set $\rho\left( R_{\rm in}\right)=1.6\times 10^{-20}\,\mathrm{g
  \ cm}^{-3}$, while the exponents for the distribution of the
temperature and rest-mass density are $k=1$ and $l=2$.

\section{Results}
\label{results}
\subsection{Relativistic Hydrodynamics} 
\label{hydro}

\begin{figure*}[h!]
\resizebox{\hsize}{!}{\includegraphics{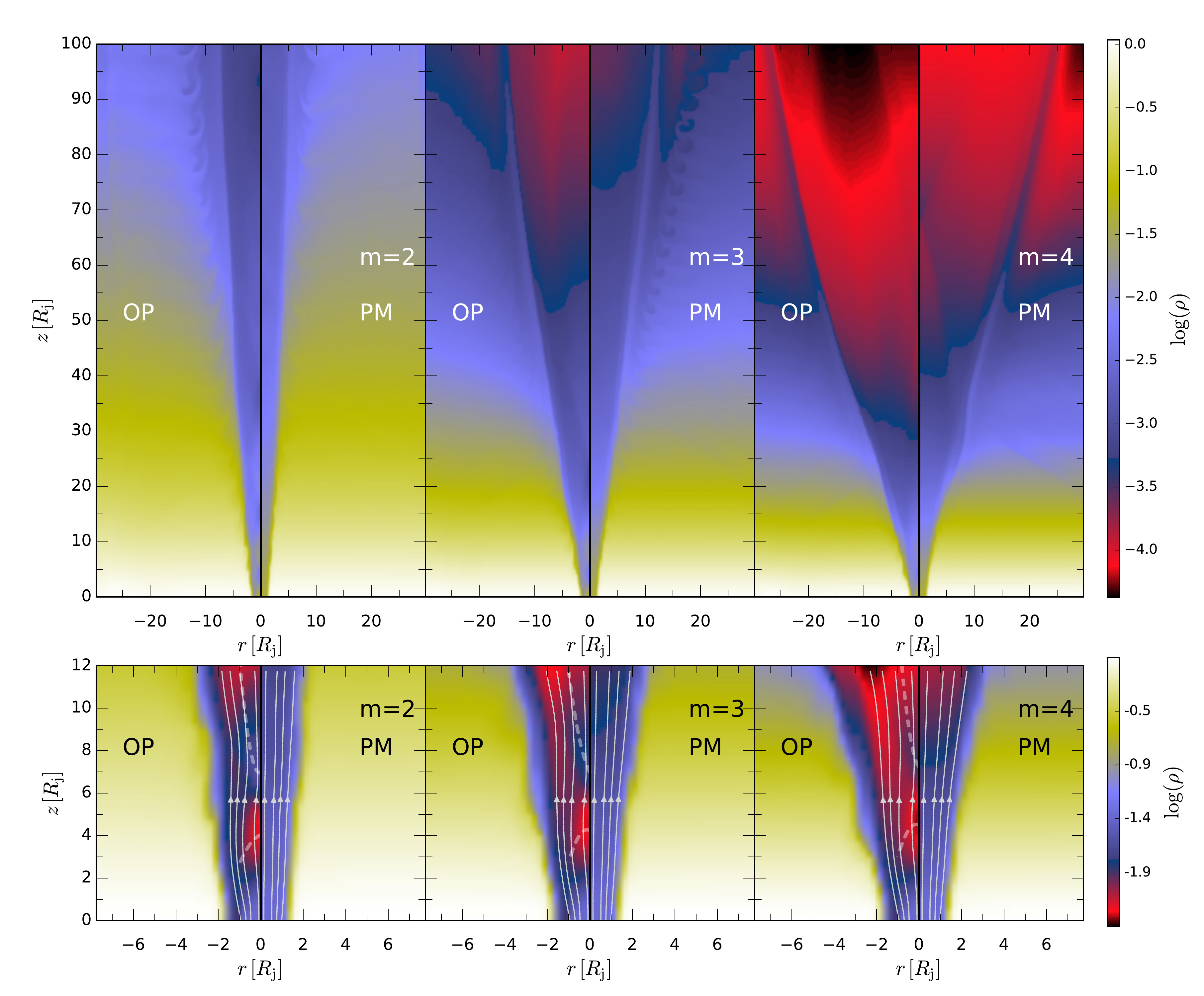}} 
\caption{Stationary results for the jet simulations. The panels show the
  2D distribution of the rest-mass density for different ambient medium
  configurations as indicated by the exponent $m$. In each panel the left
  part ($-30 < r/R_{\rm j} <0$) corresponds to the OP jet and the right
  part ($0 < y/R_{\rm j} < 30$) to the PM jet. The upper row spans the
  entire simulation grid whereas the bottom row shows a magnified view of
  the nozzle region ($-7 < z/R_{\rm j} < 7$). The white lines in the
  bottom row show stream lines visualising the direction of the flow and
  the bold dashed lines correspond to the inward travelling and reflected
  shock wave.}
\label{RHDres} 
\end{figure*}

In Fig.~\ref{RHDres} we show the 2D distribution of the rest-mass density
of the stationary jets from our (RHD) simulations. Each panel corresponds
to a different ambient medium configuration as characterised by the
exponent $m$ [see Eq.~(\ref{pamb})]. The left-half of each panel ($-30 <
r/R_{\rm j} <0$) shows the distribution for the OP jet and the right-half
of each panel ($0 < r/R_{\rm j} <30$) for the PM jet. The upper row
presents the entire simulation grid and the lower row a magnified view of
the nozzle region of the jets ($0 < z/R_{\rm j} < 7$).

In the case of OP jets, the initial pressure mismatch at the nozzle led
to the generation of two shock waves: one travelling radially outward and
the other propagating inwards in the direction of the jet axis. Between
these two shocks the jet radially expanded until a pressure equilibrium
with the ambient medium was reached. This equilibrium is first obtained
at the jet-ambient medium boundary and leads to an inward travelling wave
(indicated by the bold dashed lines in the bottom row of
Fig. \ref{RHDres}). The radial expansion of the jet is stopped as soon as
the expansion regions are crossed by the inward travelling sound
waves. Since these waves propagate with the local sound speed, the
outer regions in the jet already stop expanding while the inner regions
continue to expand (further decreasing the pressure and rest-mass
density). The innermost region in the jet stops expanding as soon as the
sound wave reaches the jet axis. At this position, a recollimation shock
is formed, which can be regarded as a new jet nozzle (local maximum in
pressure and rest-mass density) and so the flow expands again \citep[see
  \eg][]{1988ApJ...334..539D, 1991MNRAS.250..581F, Mizuno2015}. The
recollimation shocks are clearly visible at $z\sim6\,R_{\rm j}$ in the
upper-half of the panels in the left column of Fig.~\ref{RHDres}. At the
position of the recollimation shock, a new radially outward travelling
wave is formed and the jet expands again.

If the gradient in the ambient medium at the position of the first
recollimation shock ($z\sim6\,R_{\rm j}$) is steep enough, the sound
waves reflected from the boundary between the jet and the ambient medium
(generated after reaching radial pressure equilibrium) cannot reach
the jet axis and therefore no additional recollimation shock is formed
and the jet continues to expand \citep[for detailed discussion
  see][]{2015MNRAS.452.1089P}. With our settings for the ambient medium
this behaviour is obtained for models with $m>2$. In these cases, even
for the OP jets no additional recollimation shocks are formed beyond
$z_{\rm c}\sim6\,R_{\rm j}$ and a conical jet shape is obtained further
downstream.

In contrast to the OP jets, in PM jets there is a radial-pressure
equilibrium at the nozzle. Therefore, only minor variations were created
within the jet and the pressure and rest-mass density were continuously
decreasing (except for $m=4$ which triggered a shock at
$z\sim60\,R_{\rm j}$). The opening angle of these jets depends on the
gradient in the ambient medium: the steeper the gradient the larger the
opening angle of the jet. This behaviour is visible in the first column
of Fig.~\ref{RHDres}.

\begin{figure}[h!]
\resizebox{\hsize}{!}{\includegraphics{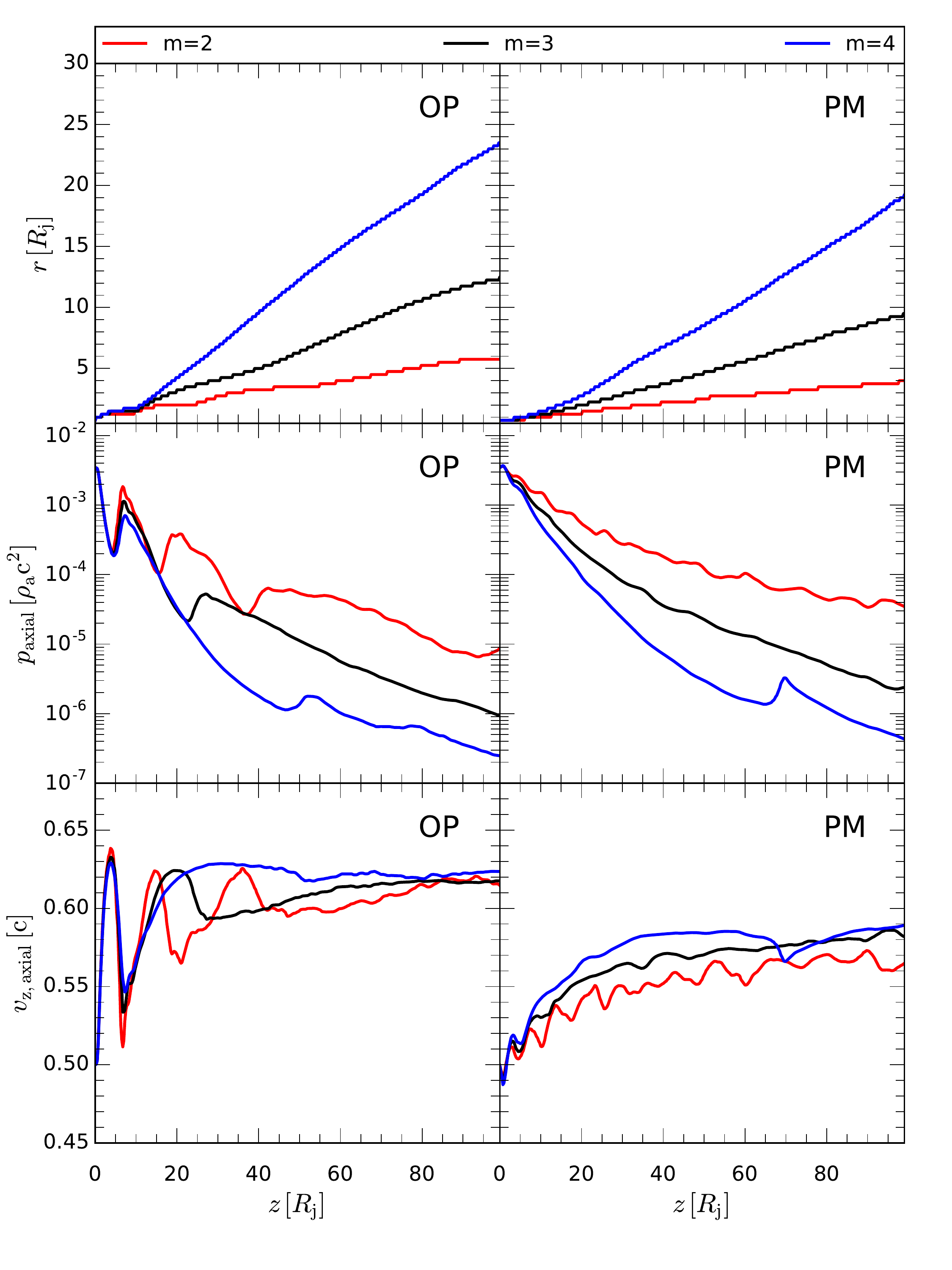}}
\caption{Variation of the jet parameters along the jet axis. The left
  column corresponds to OP jets and right one to PM jets. From top to
  bottom: jet radius, axial pressure and axial velocity in the
  $z$-direction (see text for details).}
\label{RHDcuts} 
\end{figure}

A detailed picture of the influence of the pressure ratio at the jet
nozzle, $d_k$, and the configuration of the ambient medium on the shape
of the jet and its fluid parameters can be obtained by analysing their
evolution along the jet axis. The variation of the normalised jet radius
$r/R_{\rm j}$, of the axial pressure $p_{\rm axial}$, and of the axial
velocity $v_{\rm axial}$, for the different jet and ambient medium
models is shown in Fig. \ref{RHDcuts}. The largest variations in the jet
parameters occur at small scales (\ie for $z \lesssim 20\,R_{\rm j}$)
and for different jet models, \ie pressure matched or
over-pressured. Within a jet model the modification of the gradient in
the ambient medium leads to variations, but not to a different
evolution of the parameters along the jet axis.

In the case of PM jets, stronger gradients in the ambient medium led to
larger jet radii, $r$, faster velocities, $v_{\rm axial}$, and smaller
axial pressures, $p_{\rm axial}$. This behaviour can be understood in the
following way: The radial position where a pressure equilibrium between
the jet and the ambient medium is reached (given the same initial
pressure at the jet nozzle) shifts downstream with the gradient in the
ambient medium. Since the jet is in pressure equilibrium with the ambient
medium, the axial pressure in the jet is equal to the pressure in the
ambient medium, so that the steeper the gradient in the ambient medium the
stronger the decrease in the jet pressure. According to the conservation
equations of hydrodynamics, a hot supersonic flow accelerates if its
cross section increases \citep{Rezzolla_book:2013}; thus a PM jet
embedded in an ambient medium with larger gradients (hence larger jet
radii) develops higher velocities as the same jet surrounded by an
ambient medium with smaller gradients (see bottom panel of
Fig. \ref{RHDcuts}).

In contrast to the smooth evolution of the jet parameters ($r$,
$p_{\rm axial}$, and $v_{\rm z,axial}$) for the PM jet, large
variations are present in the parameters for OP jets (see
Fig. \ref{RHDcuts}). As already mentioned the initial pressure-mismatch
at the jet nozzle leads to an opening of the jet and as soon as a radial
pressure equilibrium was reached a radially inward travelling wave was
formed. The larger the gradient in the ambient medium, the larger the jet
radii where the pressure equilibrium was obtained became. Due to the
finite sound speed, the crossing time of this wave increased with jet
radius. The inner parts of the jet stopped expanding as soon as the sound
wave crossed the jet axis. Therefore, the pressure decreased for larger
jet radii \ie a steeper gradient in the ambient medium (compare solid red
and blue lines in the middle panel of Fig.~\ref{RHDcuts}).

After the sound wave crossed the jet axis, the pressure increased to
values comparable with the pressure in the ambient medium (see middle
panel of Fig.~\ref{RHDcuts} at $z\sim8\,R_{\rm j}$. The difference in
position of the pressure jump and its magnitude reflects the difference
in the sound crossing time and the gradient in the ambient medium. Since
the OP jets exhibit larger jet radii than PM jets, the fluid is
accelerated to higher velocities, $v_{\rm z,axial}$ at $z\sim5\,R_{\rm
  j}$. The formation of the recollimation shock at $z\sim6\,R_{\rm j}$
leads to a sharp decrease in the axial velocity $v_{\rm z,axial}$, and
the expansion of the fluid after crossing the recollimation shock leads
to an increase in the axial velocity. After $z>10\,\mathrm{R_j}$, the
rapid decrease in the ambient-medium pressure induces an expansion of the
jet, which is accompanied by a decrease in the jet pressure,
$p_{\rm axial}$, and a slight increase in the axial jet velocity. In
the case of the smallest gradients in the ambient medium ($m=2$)
additional recollimation shocks are formed at $z\sim20$ and at $z\sim40$
(see the solid red curve in Fig. \ref{RHDcuts}).\\

\subsection{Emission} 
\label{emission}

For the calculation of the emission we follow the methods described in
\citet{1997ApJ...482L..33G} and \citet{Mimica:2009de}. For completeness
we provide here the basic equations and underlying assumptions for the
calculation of non-thermal emission. We assumed a power law distribution
of relativistic electrons

\begin{equation}
n\left(\gamma_{\rm e}\right)=n_0\left(\frac{\gamma_{\rm e}}{\gamma_{\rm e,\,min}}\right)^{-s}
\quad \mathrm{for\,\,\,}
\gamma_{\rm e,\,min}\leq\gamma_{\rm e}\leq\gamma_{\rm e,\, max} \,,
\end{equation}
where $n_0$ is a normalisation coefficient, $\gamma$ is the electron
Lorentz factor, $\gamma_{\rm min}$ and $\gamma_{\rm max}$ are the lower
and upper electron Lorentz factors and $s$ is the spectral slope. In
order to construct the distribution of non-thermal particles from the
thermal population we assumed that the number density of relativistic
particles is a fraction $\zeta_{\rm e}$ of the thermal particles
\begin{equation}
\int_{\gamma_{\rm e,\,min}}^{\gamma_{\rm e,\,max}}n\left(\gamma_{\rm e}\right) d \gamma=
\zeta_{\rm e}\frac{\rho}{m_{\rm p}} \,,
\label{nnumber}
\end{equation}
where $m_{\rm p}$ is the proton mass. The second assumption relates
the energy in the relativistic particles to the energy in the thermal
particles via the parameter $\epsilon_{\rm e}$ as follows
 \begin{equation}
\int_{\gamma_{\rm e,\, min}}^{\gamma_{\rm e\,max}}n\left(\gamma_{\rm e}\right)\gamma_{\rm e}
m_{\rm e} c^2  d \gamma =
\epsilon_{\rm e}\frac{p}{\hat{\gamma}-1},
\label{nenergy}
\end{equation}
where $m_{\rm e}$ is the electron mass. After performing the integration
in Eq.~(\ref{nnumber}) and Eq.~(\ref{nenergy}) a relation for lower
electron Lorentz factor, $\gamma_{\rm e,\, min}$, may be derived as
%\[
\begin{equation}
\gamma_{\rm e,\,min}=\left\{ \begin{array}{ll} 
\displaystyle
\frac{p}{\rho} \frac{m_{\rm p}}{m_{\rm e}
  c^2}\frac{(s-2)}{(s-1)(\hat{\gamma}-1)}\frac{\epsilon_{\rm e}}{\zeta_{\rm e}}
& \textrm{if }s>2 \,, \vspace*{3mm}\\
\displaystyle
\left[\frac{p}{\rho} \frac{m_{\rm p}}{m_{\rm e}
    c^2}\frac{(2-s)}{(s-1)(\hat{\gamma}-1)}\frac{\epsilon_{\rm e}}{\zeta_{\rm e}}
  \gamma_{\rm e,\,max}^{s-2}\right]^{1/(s-1)} & \textrm{if }1<s<2
\,, \vspace*{3mm}\\
\displaystyle
\frac{p}{\rho}\frac{\epsilon_{\rm e}}{\zeta_{\rm e}}
\frac{m_{\rm p}}{m_{\rm e} c^2(\hat{\gamma}-1)} \left[
  \ln\left(\frac{\gamma_{\rm e,\,max}}{\gamma_{\rm e,\,min}}\right)
  \right]^{-1} & \textrm{if } s=2 \,.
\label{gmineq1}
\end{array}\right. 
\end{equation}
%\]
%
For the upper electron Lorentz factor we used a constant fraction of the
lower electron Lorentz factor, namely 
\begin{equation}
\gamma_{\rm e,\,max}=\epsilon_\gamma\,\gamma_{\rm e,\,min} \,.
\label{gmax}
\end{equation}
The normalisation coefficient of the non-thermal particle distribution
can be obtained by performing the integral in Eq.~(\ref{nnumber}) within
the boundaries given by $\gamma_{\rm e,\, min}$ and $\gamma_{\rm
  e,\,max}$, yielding
\begin{equation}
n_0=\frac{\epsilon_{\rm e} p (s-2)}{\left(\hat{\gamma}-1
  \right)\gamma_{\rm  e,\,min}^2 m_e c^2}\left[ 1-\left(\frac{\gamma_{\rm e,\,
      max}}{\gamma_{\rm e,\,min}}\right)^{2-s}\right]^{-1} \,.
\end{equation}
Since the information on the magnetic field cannot be obtained from our
purely hydrodynamical numerical simulations, we introduce it through its
fraction $\epsilon_{\rm B}$ of the equipartition magnetic field, namely,
we set 
\begin{equation}
B=\sqrt{8\pi\epsilon_{\rm B}\frac{p}{\hat{\gamma}-1}}\,.
\end{equation}
Following \citet{Mimica:2009de}, the emission and absorption coefficients
for the non-thermal emission, respectively $\epsilon_{\nu,\mathrm{nt}}$
and $\alpha_{\nu,\mathrm{nt}}$, may be written as\footnote{We used
  unprimed variables for the co-moving frame and primed variables for the
  observer frame.}
\begin{eqnarray}
\epsilon_{\nu,\mathrm{nt}} &=& \mathcal{C} \,
n_0\gamma_{\rm e,\, min}^{s}
\int_{\gamma_{\rm e,\,min}}^{\gamma_{\rm e,\, max}}
\gamma_{\rm e}^{-s}H\left(\frac{\nu}{\nu_0\gamma_{\rm e}^2}\right) \,  d \gamma_{\rm e}\,, \\
\alpha_{\nu,\mathrm{nt}} &=& \mathcal{C} \,
n_0\frac{p+2}{m_{\mathrm{e}}\nu^2}
\gamma_{\rm e,\,min}^{s}\int_{\gamma_{\rm e,\,min}}^{\gamma_{\rm e,\,max}}
\gamma_{\rm e}^{-(s+1)}H\left(\frac{\nu}{\nu_0\gamma_{\rm e}^2}\right) \,
 d \gamma_{\rm e} \,,
\end{eqnarray}
where the characteristic frequency, $\nu_0=3\mathrm{e}B\sin{\theta}/(4\pi
m_{\rm e}c)$ with $\theta$ the angle between the direction to the
observer and the direction of the magnetic field,
$\mathcal{C}=\sqrt{3}\mathrm{e}^3 B\sin{\theta}/\left(8\pi m_{\rm e}
c^2\right)$, and $\mathrm{e}$ is the electron charge. 

The value of $H(\xi)$ depends on the orientation of the magnetic field:
If the magnetic field is ordered, $H(\xi)=F(\xi)$:
\begin{equation}
F(\xi)=x\int_x^\infty d\xi K_{5/3}(\xi)\,,
\end{equation}
where $K_{5/3}$ is the modified Bessel function of the second kind of
order $5/3$; in the case of a random magnetic field $H(\xi)=R(\xi)$, so that
\begin{equation}
R(\xi)=\frac{1}{2}\int_0^\pi d\alpha \sin^2{\alpha}
F\left(\frac{x}{\sin\alpha}\right)\,,
\end{equation}
with $\alpha$ the angle between the comoving magnetic field and the line
of sight. For the functions $F(\xi)$ and $R(\xi)$ we used the
approximations given by \cite{Crusius:1986p3163} and
\cite{Joshi:2011p2764}, namely

\begin{eqnarray}
&&F(\xi)=1.800151957\,\xi^{0.304526404}\,{\rm e}^{-\xi} \,,
\phantom{\hskip 2.0cm \textrm{for }\xi<0.01}\\
&&R(\xi)=
\left\{\begin{array}{ll}
1.4980728\,\xi^{1/3}\,, & \qquad \textrm{for } \xi<0.01\\
\\
1.08895\,\xi^{0.20949}\,{\rm e}^{-\xi} - 
& \qquad \textrm{for } \xi>0.01 
\\
\quad 0.00235861\,\xi^{-0.79051}\,{\rm e}^{-\xi}\,, 
\end{array}
\right.
\label{frapprox}
\end{eqnarray}

\begin{figure}[h!]
\resizebox{\hsize}{!}{\includegraphics{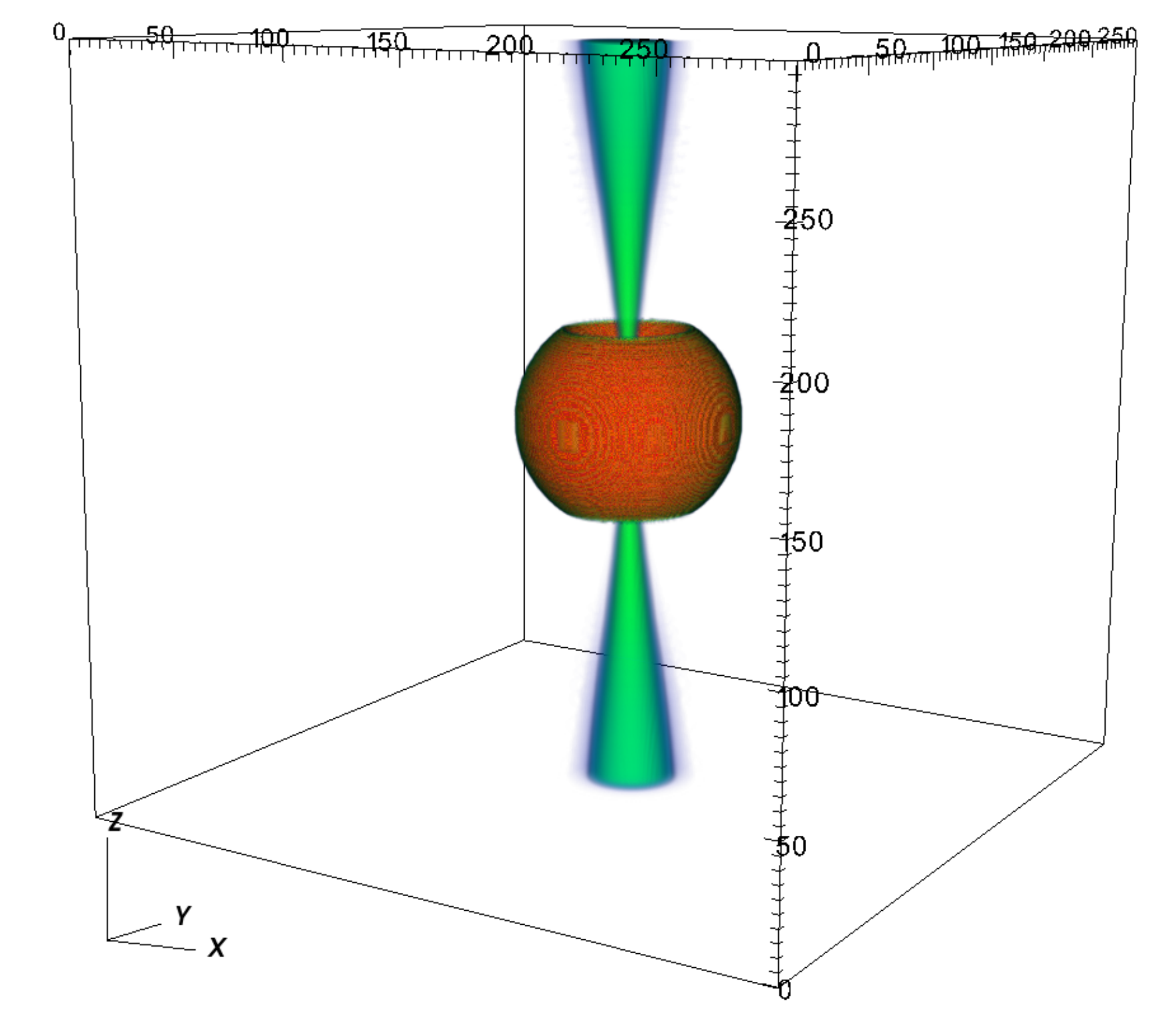}} 
\caption{Example of the 3D geometry of the jet-torus model}
\label{3dtracer} 
\end{figure}
In order to compute the observed emission, we transformed the
coefficients of emission $\epsilon_{\nu,nt}$ and absorption
$\alpha_{\nu,nt}$, into the observer's frame and corrected for
cosmological effects
\begin{eqnarray}
\epsilon^\prime_{\nu\prime,\mathrm{nt}}&=&D^{2+(p-1)/2}
\left(1+z\right)^{-(p-1)/2}\epsilon_{\nu^\prime,\mathrm{nt}} \,, \\
\alpha^\prime_{\nu^\prime,\mathrm{nt}}&=&D^{(p-1)/2+5/2}
\left(1+z\right)^{-(p-1)/2-5/2}\alpha_{\nu^\prime,\mathrm{nt}}\,,
\end{eqnarray}
where the Doppler factor is defined as $D:={\Gamma^{-1}\left(1 - v_{\rm
    b} \cos\vartheta \right)}^{-1}$ and $z$ denotes the redshift.

In addition to the non-thermal component, the emission includes also the
contribution from a dusty torus, so that the thermal absorption
coefficient $\alpha_{\nu,\mathrm{th}}$ is given by
\begin{equation}
\alpha_{\nu,\mathrm{th}}=3.7\times 10^{8}\,
T^{-1/2}Z^2\frac{\rho^2}{m_{\rm p}^2}\nu^{-3}
\left(1-\mathrm{e}^{h\nu/k_{\rm b}T}\right)\bar{g}_{\mathrm{ff}}\,,
\end{equation}
where $Z=\mu_{\rm i}/\mu_{\rm e}$ with $\mu_{\rm e}=2/(1 + X)$ and
$\mu_{\rm i}=4/(1 + 3X)$ where $X$ corresponds to the abundance of
hydrogen (set here to be $X=0.71$) and $\bar{g}_{\mathrm{ff}}$ is the
velocity-averaged gaunt factor \citep[see,
  \eg][]{1986rpa..book.....R}. For the gaunt factor we used the values
tabulated by \citet{2014MNRAS.444..420V} and applied a bi-linear
interpolation in $\log u$--$\log \gamma^2$ space, where
$u=({h\nu})/({k_{\rm b}T})$ and $\gamma^2=({Z^2\mathrm{Ry}})/({k_{\rm
    b}T})$, and $\mathrm{Ry}$ is the Rydberg energy.

The total intensity, $I_\nu$, was a mixture of the non-thermal
emission/absorption and thermal absorption (if the ray propagated through
the dusty torus) and was computed via the transport equation as
\begin{equation}
\frac{ d I_\nu}{ d s}=\epsilon_{\nu,\mathrm{nt}} -
\left(\alpha_{\nu,\mathrm{nt}} + 
\alpha_{\nu,\mathrm{th}}\right)I_\nu\,,
\end{equation}
where $d s$ is the path length along a ray. For the calculation of the
emission we computed the starting point of each ray at the position where
the optical depth $\tau=\tau_{\rm limit}$ and followed the ray to the
observer. The optical depth is defined as
\begin{equation}
\tau=\int_0^{s^\prime}\left(\alpha_{\nu,\mathrm{nt}} +
\alpha_{\nu,\mathrm{th}}\right) d s \,.
\end{equation}
The observed flux density can be computed from the intensity according to
\begin{equation}
S_\nu=\left( \frac{1+z}{D_{\mathrm{L}}^2} \right) \Delta x \, \Delta y \,
I_\nu \,,
\label{obsflux}
\end{equation}
where $D_L$ is the luminosity distance, $\Delta x$ and $\Delta y$
represent the resolution of the detector's frame. The luminosity distance
is computed via the equation provided by \citet{1999ApJS..120...49P},
where we took $\Omega_{\rm m}=0.27$ and a Hubble constant of
$H_0=71\,\mathrm{km \, s^{-1} Mpc^{-1}}$. The final jet-torus geometry at
a viewing angle of $\vartheta=0^\circ$ used for the calculation of the
emission is shown in Fig.~\ref{3dtracer}; furthermore, hereafter we
assume $\left(D_{\rm L}=21.1\,\mathrm{Mpc}\right)$, which corresponds to
a redshift of $z=0.005$.

\subsection{Parameter space study}

In this Section we analyse the influence of the emission and torus
parameters on the observed spectrum. In Table~\ref{paraem} we provide an
overview of the parameters used and of the values for our reference
model. Quite generically, such parameters can be divided into three
classes: i) ``scaling parameters'', ii) ``emission parameters'', and iii)
``torus parameters''; we will discuss them separately below.

\begin{table}[h!]
\caption{Parameters for the emission simulations and values for the
  reference model}
\label{paraem}
\centering
\begin{tabular}{@{}l l l@{}}
\hline\hline
Symbol  & Value &  Description\\
\hline
$d_{\rm k}$					  &1.0											& \tiny pressure mismatch at nozzle\\
$n$						 &1.5												& \tiny pressure gradient in ambient medium\\
$m$						&2.0												& \tiny pressure gradient in ambient medium\\
$z$						  &0.005											& \tiny redshift\\
$R_{\rm j}$  				         &$3\times10^{16}$ cm							&  \tiny jet radius	\\
$\rho_{\rm a} $				 &$1.67\times 10^{-21}\,\mathrm{g \, cm^{-3}}$	       		& \tiny ambient medium density	\\
$\epsilon_{\rm B}$			 &0.1												& \tiny equipartition ratio\\
$\epsilon_{\rm e}$ 		        &0.3												& \tiny thermal to non-thermal energy ratio\\
$\zeta_{\rm e}$ 				 &1.0											& \tiny thermal to non-thermal  number density ratio\\
$\epsilon_\gamma$		 &1000											& \tiny ratio between e$^-$ Lorentz factors\\
$s$						 &2.2											& \tiny spectral index\\
$\vartheta$				 &80$^\circ$ 										& \tiny viewing angle\\
$L_{\rm AGN}$		&$1\times10^{43}\,\mathrm{erg \,s^{-1}}$			& \tiny bolometric luminosity\\
$R_{\rm out}$		&$3\times10^{17}\,\mathrm{cm}$					& \tiny torus outer radius\\
$\theta$					&50$^\circ$ 										& \tiny torus thickness\\
$\rho\left(R_{\rm in}\right)$	 &$1.67\times 10^{-19}\,\mathrm{g \, cm^{-3}}$		& \tiny torus density at  $R_{\rm min}$\\
$T_{\rm sub}$			 &1500 K									& \tiny dust sublimation temperature\\
$k$, $l$					&1, 2					 					&\tiny exponents for $\rho$ and $T$ distributions\\
\hline
\end{tabular}
\end{table}

In order to cover a wide frequency range, \ie $10^9~\mathrm{Hz} < \nu <
10^{12}~\mathrm{Hz}$, we used a logarithmic grid with 100 frequency
bins. The ray-tracing was performed using a Delaunay triangulation on the
axisymmetric RHD simulations with $(r,\,z)$ coordinates to a 3D Cartesian
grid $(x,\,y,\,z)$ with a typical resolution of $300^3$ cells (see
Fig.~\ref{3dtracer} for a 3D representation of the jet-torus system).

\subsubsection{Single-dish spectra}
\label{singledishpara}

\begin{figure}[h!]
\resizebox{\hsize}{!}{\includegraphics{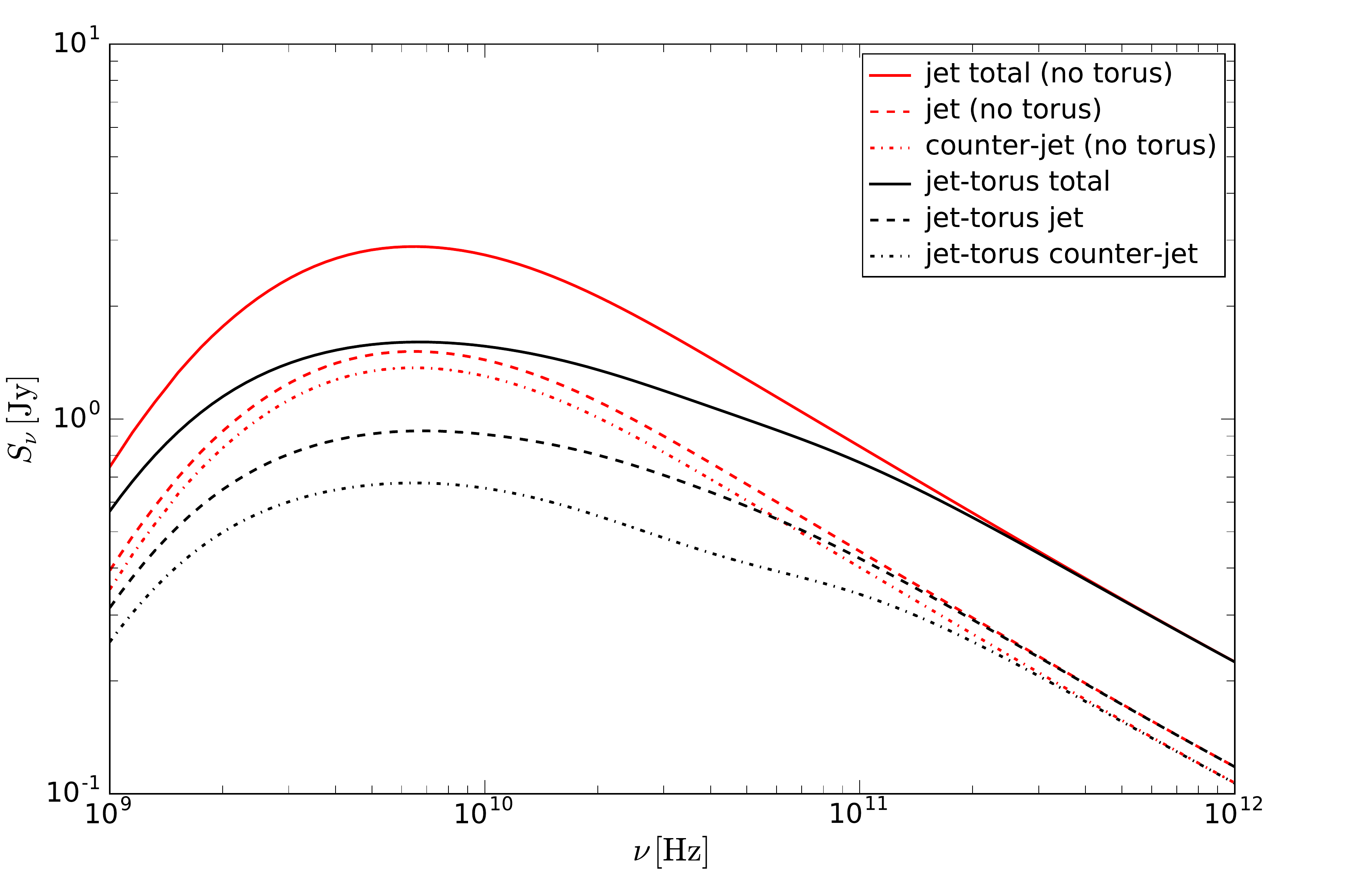}} 
\caption{Single-dish spectra for a simulation including a torus (black)
  and without a torus (red).}
\label{torusnotorus} 
\end{figure}

To demonstrate the influence of the torus on the spectrum we computed the
emission for the reference model with and without a torus. The result of
these calculations is presented in Fig.~\ref{torusnotorus}, where the
black lines indicate simulations including a torus, while red lines are
used for simulations without a torus. Furthermore, solid lines refer to the
total flux density, while dashed and dotted lines correspond to the
emission from the jet and from the counter-jet, respectively.

Not surprisingly, in the low-frequency regime, \ie for $\nu
<10^{11}\,\mathrm{Hz}$, simulations with a torus show a smaller flux than
simulations without a torus, simply as a result of the increased
absorption along the line of sight. At higher frequencies, the flux
densities of both simulations are very similar since the torus becomes
optically thin at these frequencies. The comparison between the emission
from the jet (dashed lines) and from the counter-jet (dotted lines) shows
that the counter-jet is more affected by the absorption of the torus than
the jet. This is to be contrasted with the results obtained in the case
of simulations of jets with a torus, where the emission converges around
$\nu\sim10^{11}\,\mathrm{Hz}$ to the flux density of the jet without
torus. A similar behaviour was obtained for the counter-jet but at a
higher frequency, \ie $\nu\sim3\times 10^{11}\,\mathrm{Hz}$.

\begin{figure}[h!]
\resizebox{\hsize}{!}{\includegraphics{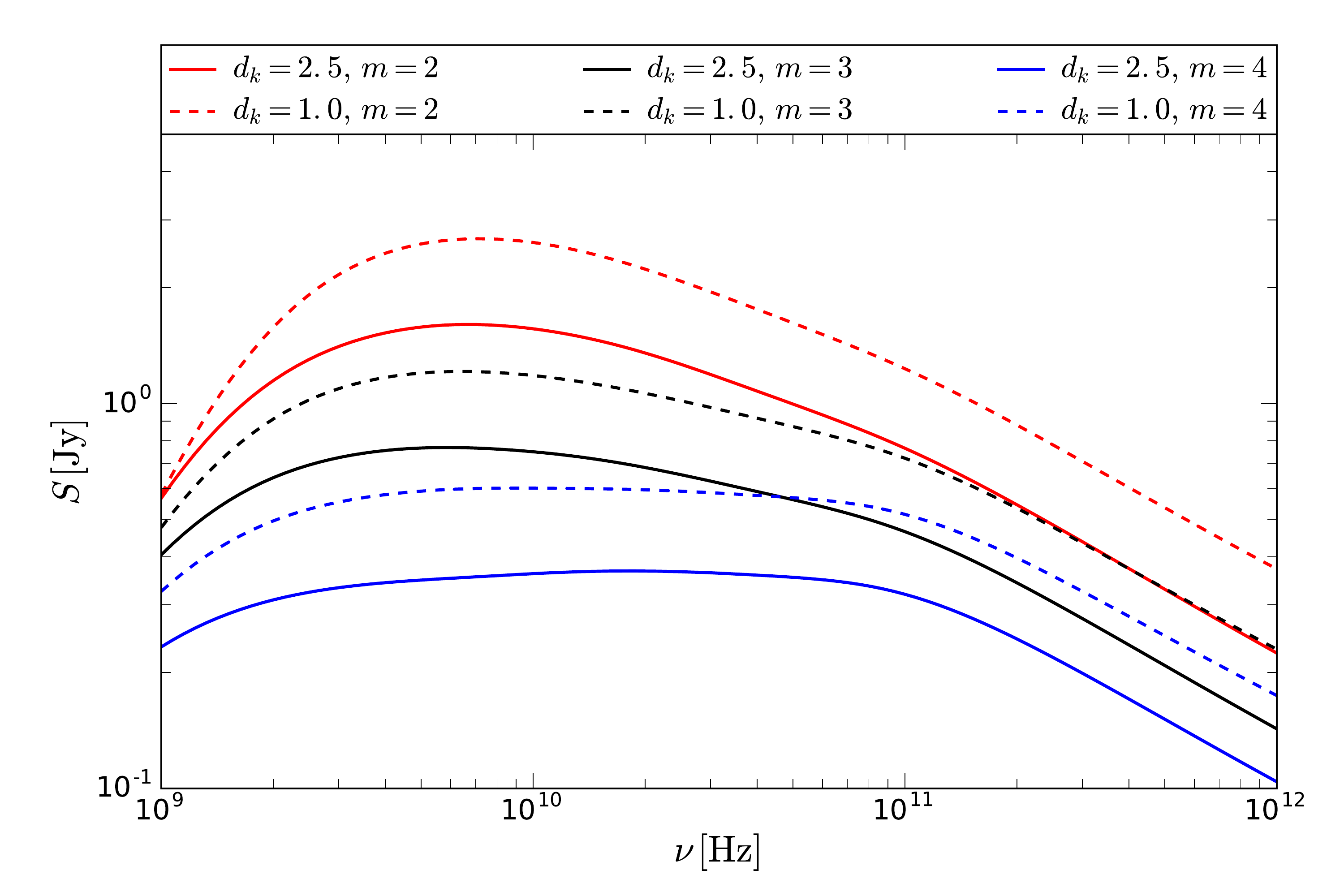}} 
\caption{Single-dish spectra for the different jet and ambient medium
  models.}
\label{jetparaspace} 
\end{figure}

Figure \ref{jetparaspace} shows the single-dish spectrum for the
different jet-torus and ambient medium models (see
Table~\ref{jpara}). Within the same ambient medium, the PM jets ($d_{\rm
  k}=1$, dashed lines in Fig.~\ref{jetparaspace}) exhibits higher flux
densities than the OP jets ($d_{\rm k}=2.5$, solid lines in
Fig.~\ref{jetparaspace}). Furthermore, OP jets show a flatter spectrum
than PM jets within $10^9\,\mathrm{Hz}\leq\nu\leq10^{11}\,\mathrm{Hz}$.
The general trend could be described as follows: the steeper the gradient
in the ambient medium, the flatter the spectrum. The change in the
spectral slope at $\nu\sim10^{11}\,\mathrm{Hz}$ is indicative of the
turnover frequency of the torus, \ie the frequency at which the torus
becomes optically thin (see also Fig.~\ref{torusnotorus}).

\begin{figure*}[t!]
\centering 
\includegraphics[width=17cm]{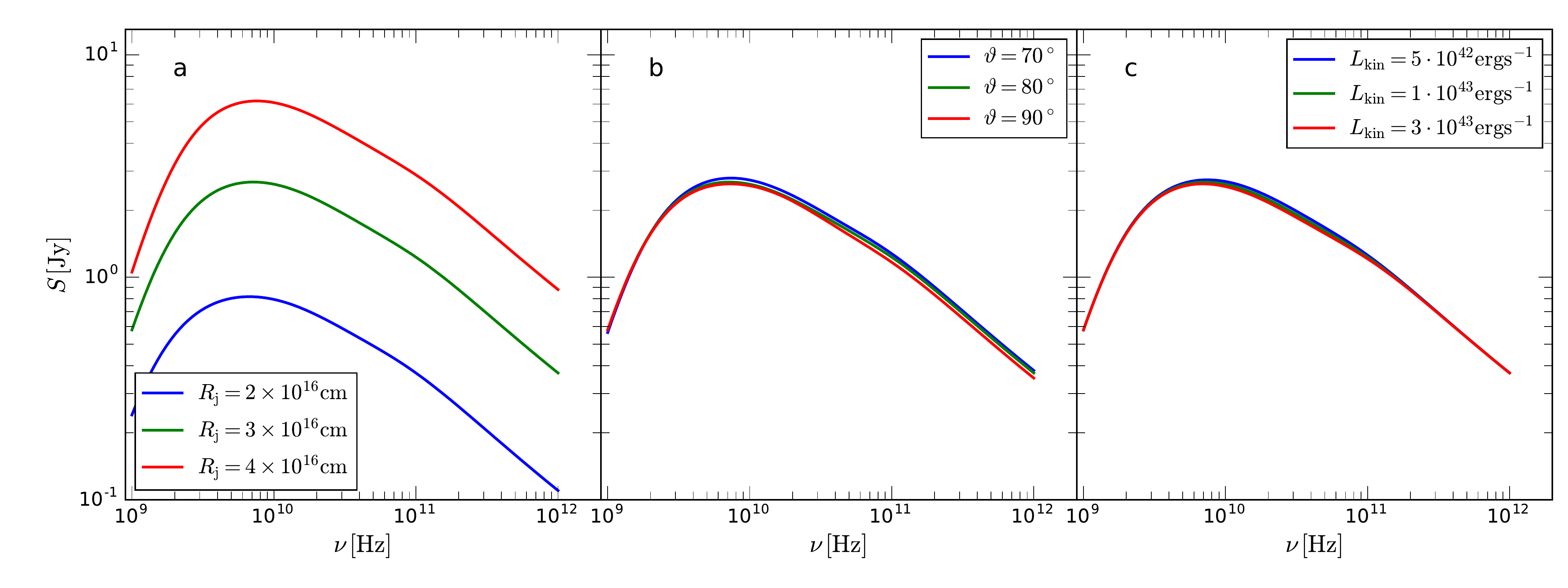} 
\caption{Influence of the scaling parameters $R_{\mathrm{j}}$,
  $\vartheta$ and $L_{\mathrm{kin}}$ on the shape of the observed
  single-dish spectrum}
\label{scalingpara}
\end{figure*}

Figure \ref{scalingpara} is meant to summarise the impact of the
different scaling parameters, $R_{\rm j},\,\vartheta,\, \mathrm{and},\,
L_{\rm kin}$, on the single-dish spectrum. More specifically, we have
computed the emission when only one of the scaling parameters is varied
while the others are kept fixed to the values listed in
Table~\ref{paraem}. The variation in the observed flux can then be
summarised as follows:

\begin{itemize}
\item
\emph{Jet radius, $R_{\rm j}$}: The jet radius acts as the scaling length
of the simulations and has the largest impact on the observed single-dish
spectrum (see panel a in Fig.~\ref{scalingpara}). The increase in flux
with $R_{\rm j}$ follows a simple scaling $\sim \Delta R_{\rm j}^2$, as
expected from Eq.~(\ref{obsflux}).

\item
\emph{Viewing angle, $\vartheta$}: Due to Doppler boosting, the observed
spectrum is shifted towards higher flux densities and higher turnover
frequencies with decreasing viewing angle. Since we use large viewing
angles for our simulations the variations in the spectrum are minor.

\item
\emph{Bolometric luminosity, $L_{\rm kin}$}: Changes in the bolometric
luminosity lead to a variation in the inner radius of the torus [\cf
  Eq.~(\ref{rinner})]. Lowering the kinetic luminosity decreases the
inner radius of the torus. Since we keep the outer radius of the torus
fixed, the total size of the torus increases with decreasing kinetic
luminosity and smaller values for rest-mass density and temperature are
obtained. Since the opacity of the torus scales like $\tau_{\rm
  torus}\propto T^{-1/2}\rho^2$ (when assuming $T$ and $\rho$ to decay as
$r^{-k}$), lower densities lead to lower opacities and a large emission
from the jet. As a result, the observed flux density increases with
decreasing kinetic luminosity.
\end{itemize}

\begin{figure*}[t!]
\centering 
\includegraphics[width=17cm]{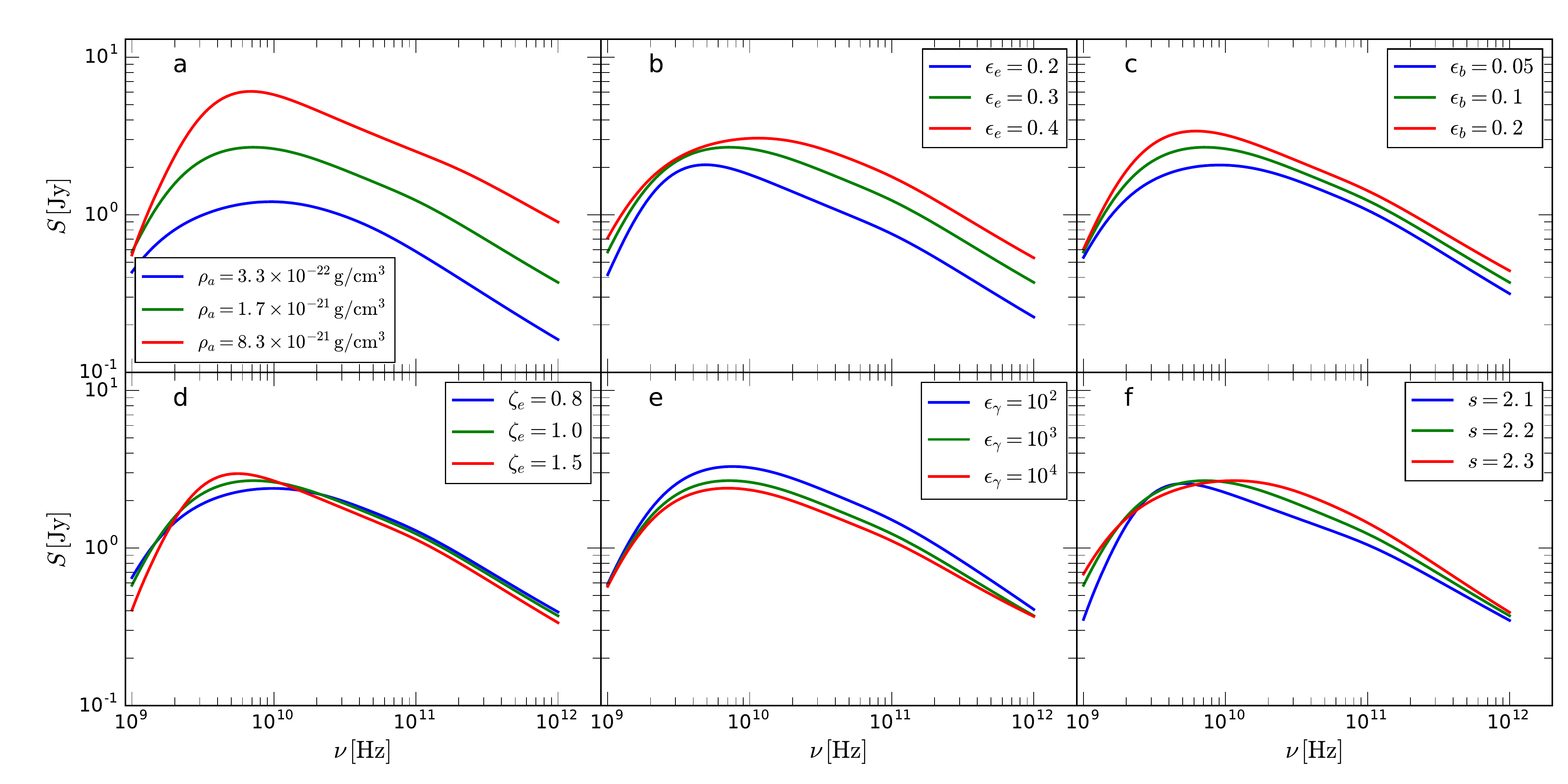} 
\caption{Influence of the emission parameters $\rho_a,\, \epsilon_e,\,
  \epsilon_b,\, \zeta_e,\, \epsilon_\gamma,\, \mathrm{and}\, s$ on the
  shape of the observed single-dish spectrum.}
\label{eeparaspace}
\end{figure*}

In Fig.~\ref{eeparaspace} we report the effect of the emission parameters
$\epsilon_{\rm e},\,\epsilon_{\rm b},\, \zeta_{\rm e},\, \epsilon_\gamma,\, \mathrm{and}\, s$ on the observed single-dish spectrum while keeping the
other parameters fixed. The emission simulations were performed using our
reference model (see Table~\ref{paraem}).

\begin{itemize}
\item
\emph{Ambient medium rest-mass density, $\rho_{\rm a}$}: A variation in
the ambient medium rest-mass density $\rho_{\rm a}$ leads to large
changes in the shape and the total flux density of the spectrum (see
panel a in Fig.~\ref{eeparaspace}). The single-dish spectrum steepens and
is shifted to higher turnover flux densities with increasing ambient
medium rest-mass density.

\item
\emph{Thermal to non-thermal energy ratio,$\epsilon_{\rm e}$}: If more
energy is stored in the non-thermal particles via an increase in
$\epsilon_{\rm e}$, the turnover frequency increases, while the turnover
flux density rises only slightly. The spectral index within $\nu \sim
10^{11}\,{\rm Hz}$ decreases in absolute values, thus leading to a
flatter spectrum (see panel b in Fig.~\ref{eeparaspace}).

\item
\emph{Equipartition ratio, $\epsilon_{\rm b}$}: An increase of the
equipartition fraction $\epsilon_{\rm b}$ leads to higher observed flux
densities on the single-dish spectrum and to a shift of the turnover
position towards smaller frequencies and higher flux densities (see panel
c of Fig.~\ref{eeparaspace}). Additionally, the spectrum steepens for
$\nu<10^{11}\,{\rm Hz}$.

\item
\emph{Ratio between thermal to non-thermal number densities, $\zeta_{\rm
    e}$}: Increasing the number of non-thermal particles leads to higher
observed flux densities and shifted the turnover frequency towards lower
frequencies. Additionally, the spectral slope increases (see panel d of
Fig.~\ref{eeparaspace}.

\item
\emph{Ratio between upper and lower $\mathrm{e}^{-}$-Lorentz factors,
  $\epsilon_\gamma$}: The variation of $\epsilon_\gamma$ had only a minor
effect on the single-dish spectrum (see panel e in
Fig.~\ref{eeparaspace}). An increase in $\gamma_{\rm max}$ leads to a
flattening of the spectrum at high frequencies ($\nu>10^{11}\,{\rm Hz}$)
and shifted the high-frequency cut-off in the spectrum towards higher
frequencies, which are clearly visible for $\gamma_{\rm e}=10^2$ (see
blue line in panel e).

\item
\emph{Spectral slope, $s$}: Changing the spectral slope of the
non-thermal particle distribution induced only minor variations in the
observed flux densities. However, the slope of the spectrum for
$\nu\leq10^{11}\,\mathrm{Hz}$ flattens, while for $\nu>10^{11}\,{\rm Hz}$
the spectral slope steepens with increasing $s$ (see panel f in
Fig.~\ref{eeparaspace}).
\end{itemize}

\begin{figure*}[t!]
\centering 
\includegraphics[width=17cm]{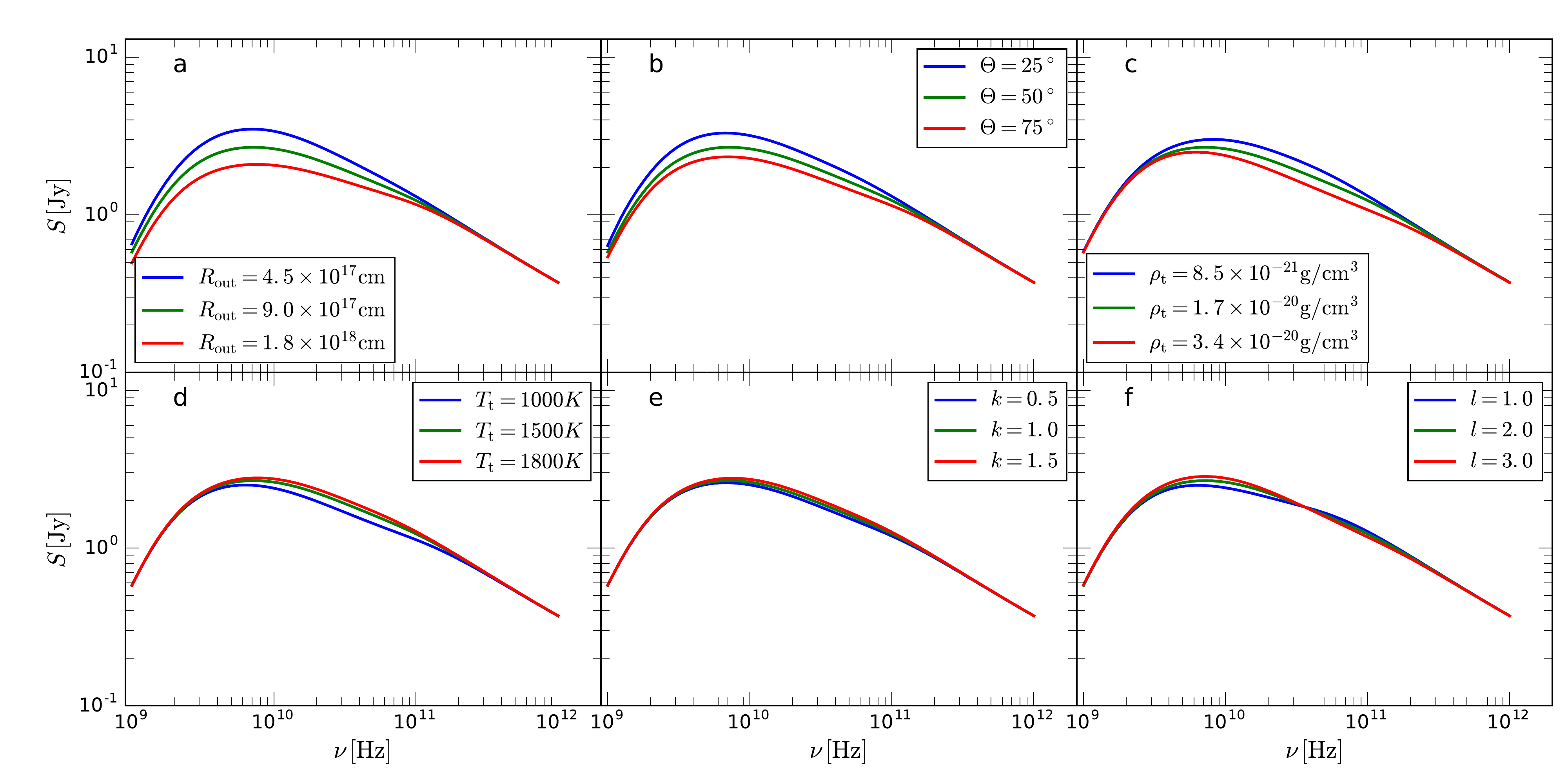} 
\caption{Influence of the torus parameter on the observed single-dish
  spectrum. Top row from left to right: torus outer radius, torus
  thickness and torus rest-mass density. Bottom row from left to right: torus
  temperature, the distribution of the torus rest-mass density, and the
  distribution of the torus temperature}
\label{toparaspace}
\end{figure*}

Since the properties of the obscuring torus strongly affect the observed
emission, we have also performed a parameter-space study varying the
torus' properties reported in Table \ref{paraem}. We recall that in our
model the torus is characterised by its geometry ($R_{\rm out}$ and
$\Theta$) and by the distribution of the rest-mass density and the
temperature [see Eqs.~(\ref{rhotorus})--(\ref{temptorus})]. The first row
in Fig.~\ref{toparaspace} summarises the changes on the single-dish
spectrum result from the torus geometry, while the second row shows the
modification in the spectral shape depending on the distribution of the
rest-mass density and temperature. Overall, the variations in the
spectrum can be summarised as

\begin{itemize}
\item
\emph{Torus geometry}: The variation of the torus geometry and its effect
on the single-dish spectrum is presented in the top row of
Fig.~\ref{toparaspace}. From left to right we show the effects of the
torus outer radius, $R_{\rm out}$, the torus thickness, $\Theta$, and the
torus rest-mass density, $\rho_{\rm t}$. Increasing the outer radius of
the torus leads to a drop in the observed emission between
$10^9\,\mathrm{Hz}\leq\nu\leq10^{11}\,\mathrm{Hz}$. The influence of the
torus thickness is presented in panel b of Fig. \ref{toparaspace}, which
shows that the observed flux density decreases as the thickness of the
torus is increased.

\item
\emph{Density distribution}: The observed flux density decreases with
increasing torus rest-mass density. At the same time, the turnover
frequency is shifted towards higher frequencies (see panel d in
Fig.~\ref{toparaspace}).

\item
\emph{Temperature distribution:} Variations in the torus' temperature
have only minor impacts on the observed flux density, so that the hotter
the torus the higher the observed flux density (see panel g in
Fig. \ref{toparaspace}).
 
\item
\textit{Radial and angular distribution of the torus rest-mass density
  and torus temperature}: The gradient in the radial direction modifies
mainly the total observed flux density, with a steeper gradient in the
radial direction leading to an increase in the flux density (see panel e
in Fig.~\ref{toparaspace}). On the other hand, a larger effect on the
shape of the spectrum comes from changing the gradient in the polar
direction (see panel f). More specifically, if the temperature
distribution is more concentrated at the edges and the rest-mass density
falls off more rapidly towards the edges of the torus (\ie if larger
values for $l$ are used), the spectrum steepens and the observed flux
density is higher than for smoother rest-mass density and temperature
distributions (\ie for smaller values for $l$).
\end{itemize}

All things considered, the parameter study on the single-dish spectrum
can be summarised as follows: The scaling parameter $R_{\rm j}$ and the
ambient medium rest-mass density $\rho_{\rm a}$ have the largest impact on the observed
single-dish flux density (see panel a in Fig. \ref{scalingpara} and
Fig. \ref{eeparaspace}). Changes in the emission parameters,
$\epsilon_{\rm e},\,\epsilon_{\rm b},\, \zeta_{\rm e},\,
\epsilon_\gamma,\, \mathrm{and}\, s$ mainly modify the shape of the
spectrum and lead only to minor variations in the flux density (see
Fig. \ref{eeparaspace}). The torus parameters show the smallest effect
and alter mainly the slope in the optically thin regime of the spectrum
(see Fig. \ref{toparaspace}).

\subsubsection{Synthetic radio maps}
\label{syntheticradio}

To facilitate comparison with VLBI observations, we computed radio maps
at different frequencies and convolved them with a 2D Gaussian beam. The
size of the beam depends on the properties of the array and on the
observing frequency. In Table~\ref{beams} we present the typical beam
sizes {(uniform weighting, full-track $(u,\,v)$ coverage for a source at
  $-8^\circ$ declination)} for the Very Long Baseline Array (VLBA) and
its sensitivity, \ie the lowest detectable thermal noise (see the EVN
calculator\footnote{http://www.evlbi.org/cgi-bin/EVNcalc} for details).
If not stated otherwise, a flux-density cut off of $5\times S_{\rm
  limit}$ was used for the synthetic radio maps.

\begin{table}[h!]
\caption{VLBA properties used for the convolution of the radio maps.}
\label{beams}
\centering
\begin{tabular}{c c c c}
\hline\hline
$\nu$ & $b_{\rm maj}$ & $b_{\rm min}$ & $S_{\rm limit}$\\ 
 $[$GHz] & [mas] & [mas] & [$\mu$J]\\
 \hline
 5      & 3.30	& 1.30	&22\\
 8	& 1.98 	& 0.81	&27\\
15	& 1.28	& 0.50	&43\\
22	& 0.86	& 0.32	&50\\
43	& 0.45	& 0.16	&93\\
86	& 0.35	& 0.06	&421\\
\hline
\end{tabular}
\end{table}

\begin{figure}[h!]
\resizebox{\hsize}{!}{\includegraphics{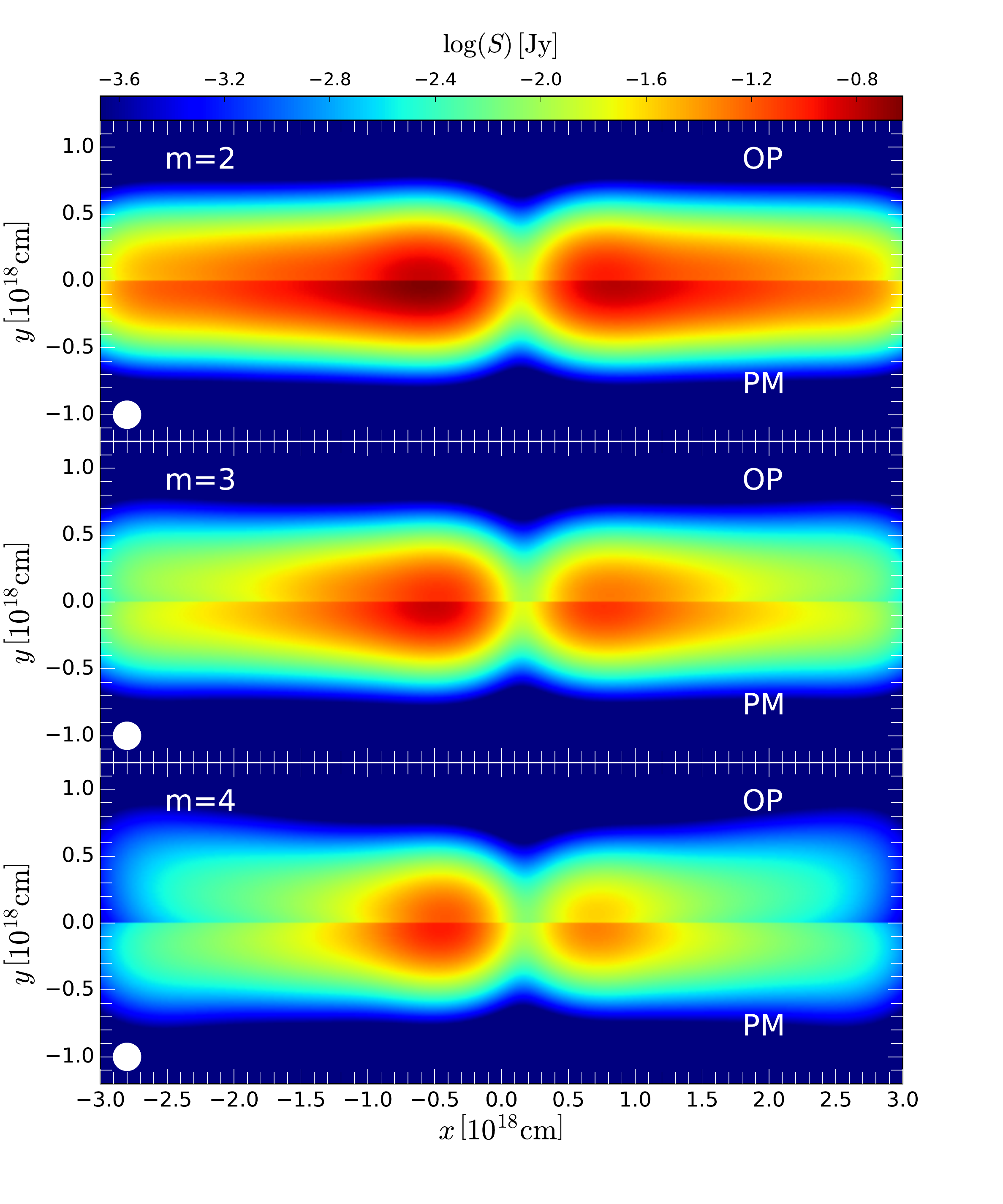}}
\caption{Convolved radio maps at $15\,{\rm GHz}$. The panels correspond
  to different gradients used in the ambient medium indicated by $n$ and
  $m$. In each panel the the upper half corresponds to an OP jet and the
  lower half to a PM jet. The convolving beam is plotted in lower left
  corner of each panel.}
\label{radiomaps15} 
\end{figure}

In Fig.~\ref{radiomaps15} we show the convolved radio maps at $15\,{\rm
  GHz}$ for different ambient medium and jet models (similar to
Fig. \ref{RHDres}). The panels show, from top to bottom, different models
for the ambient medium and in each panel the upper half (\ie $(0 - 1)
\times 10^{18}\,{\rm cm}$ corresponds to the OP jet and lower half (\ie
$(-1 - 0) \times 10^{18}\,{\rm cm}$) to the PM jet. The width of the jet
increases with the gradient in the ambient medium, from top to bottom,
from $0.5\times10^{18}\,{\rm cm}$ to $0.8\times10^{18}\,{\rm cm}$, while
the observed flux density decreases. Within a common atmosphere, the OP
jets (upper half of each panel) show a less elongated core region as
compared to the PM jets (lower half of each panel) and the axial flux
density decreases faster in OP jets than in PM jets. These effects become
more visible as the rest-mass density gradient in the ambient medium is
increased.

\begin{figure*}[t!]
\centering
\includegraphics[width=17cm]{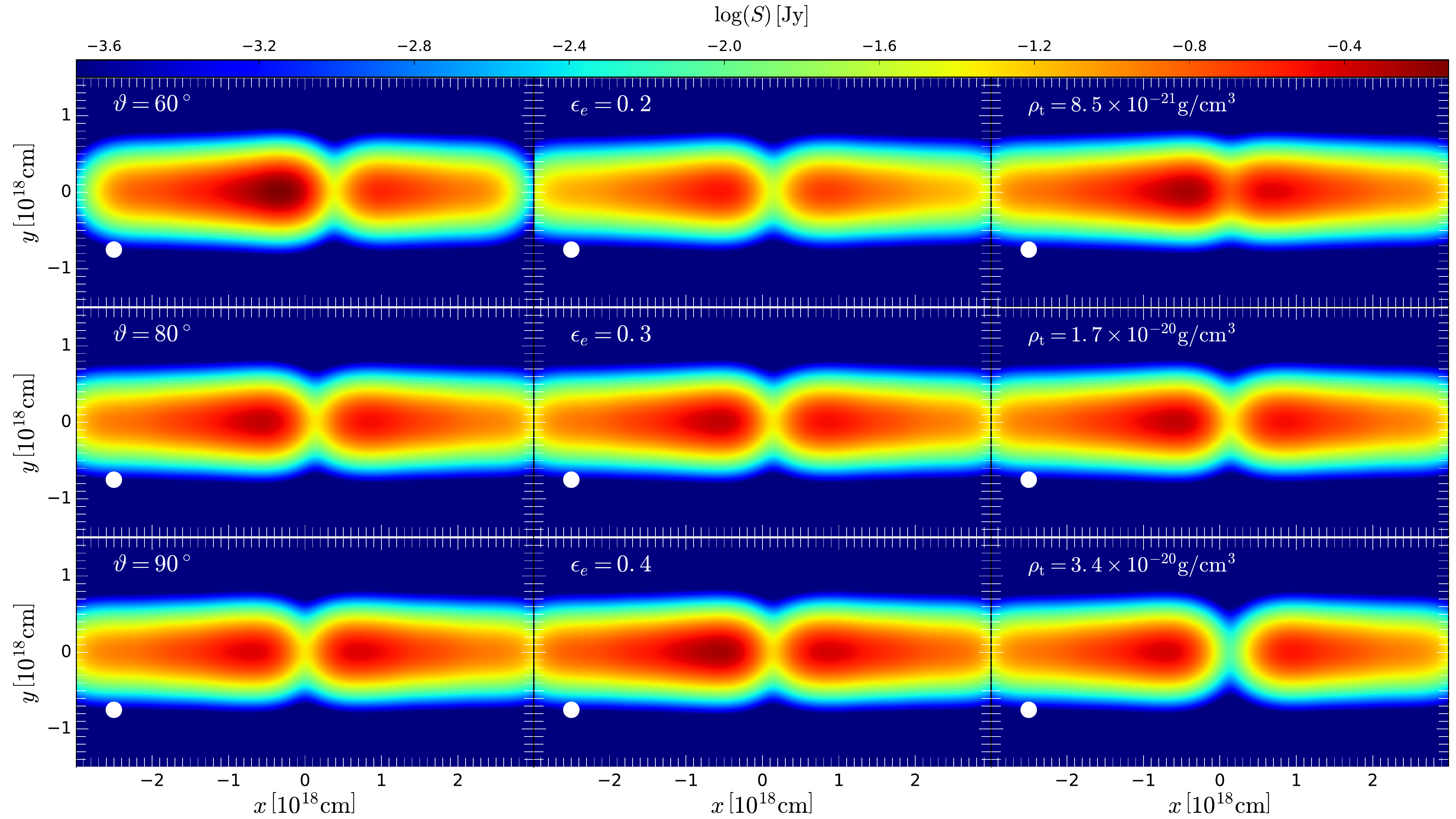}
\caption{Influence of the viewing angle (left column), the thermal to
  non-thermal energy ratio (middle column) and the torus rest-mass
  density (right column) on the $15\,{\rm GHz}$ radio maps. The
  convolving beam is plotted in the lower left corner of each panel.}
\label{radiotorus}
\end{figure*}

Analogous to the parameter study performed on the single-dish spectra
(see Sec.~\ref{singledishpara}), we have investigated the impact of the
emission parameters presented in Table~\ref{paraem} on the synthetic
radio maps. Figure~\ref{radiotorus} summarises the effects of the viewing
angle $\vartheta$, of the thermal to non-thermal energy ratio
$\epsilon_e$, and of the rest-mass density of the torus $\rho_{\rm t}$,
on the $15\,{\rm GHz}$ radio maps. For a better comparison, the reference
model is included in the second row of Fig.~\ref{radiotorus}.

At smaller viewing angles (\ie for $\vartheta=60^\circ$), the jets appear
more asymmetric due to the increased Doppler boosting and larger opacity
along the line of sight, especially for the counter-jet (see the top
panel in the first row of Fig.~\ref{radiotorus}). However, with
increasing viewing angle the asymmetries between jet and counter-jet tend
to disappear, while the total flux density in the radio map
decreases. Upon varying the non-thermal to thermal energy ratio
$\epsilon_{\rm e}$, the flux density increases and affects both jets in
the same way (see second row in Fig.~\ref{radiotorus}). On the other
hand, a change in $\epsilon_{\rm e}$ does not alter the jet structure or
the emission gap between the jet and the counter-jet.

In the third row of Fig.~\ref{radiotorus} we show the changes in the
$15\,{\rm GHz}$ radio maps due to a variation in the torus rest-mass
density, $\rho_{\rm t}$. Since we keep the torus size constant, a change
in the torus rest-mass density does not have an impact on the observed
emission at $\left| x \right |>1\times 10^{18}\,{\rm cm}$. However, the
observed gap between jet and counter-jet does change depending on the
torus rest-mass density. In particular, the gap is nearly invisible at
low densities (top panel in the third row) while it is clearly noticeable
at higher densities (see last panel in the third row of
Fig.~\ref{radiotorus}). Besides the changes in the gap between the jet
and counter-jet, a variation in the torus rest-mass densities leads to
changes in the observed flux densities within $\left| x\right | <1\times
10^{18}\,{\rm cm}$. This can be easily explained as due to the increased
opacity, which reduces the flux reaching the observer (see the decreasing
flux density at $x=0.5\times10^{18}\,{\rm cm}$).

In the analysis of VLBI observations it is common to compute spectral
index maps, where the spectral index, $\alpha$, between to adjacent
frequencies $\nu_1$ and $\nu_2$ is given by
\begin{equation}
\alpha_{\nu_1,\nu_2} := \frac{\log\left({S_{\nu_1}/{S_{\nu_2}}}\right)}
      {\log\left(\nu_1/\nu_2\right)} \,.
\end{equation}
To avoid resolution-induced artefacts in the spectral index maps, a
common beam size was used for the convolution. Furthermore, as is usually
done in observational VLBI in order to avoid over-resolution from the
high-frequency image, the beam size was set on the low-frequency map.

\begin{figure}[h!]
\resizebox{\hsize}{!}{\includegraphics{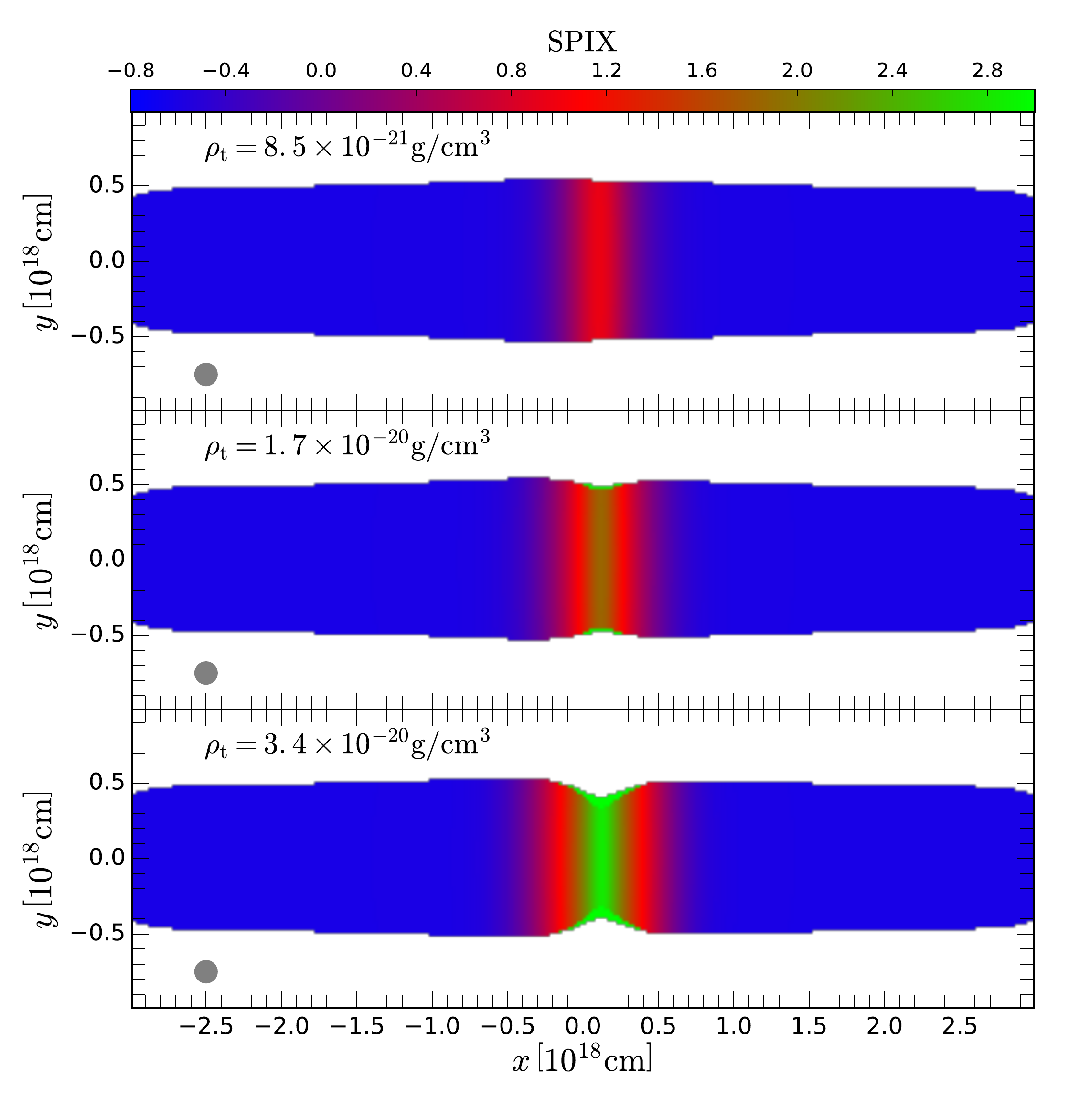}}
\caption{Spectral index maps computed between 22\,GHz and 43\,GHz for
  different torus densities. The
  convolving beam is plotted in the lower left corner of each panel.}
\label{spix1543} 
\end{figure}

In Fig.~\ref{spix1543} we present the spectral index computed between
$22\,{\rm GHz}$ and $43\,{\rm GHz}$ for different torus densities
indicated in the top left corner. The spectral index (SPIX) is found to
be negative (\ie $\alpha\sim-0.6$) for regions $|x|>0.5\times
10^{18}\,{\rm cm}$ to reflect optically thin regions, and positive (\ie
$\alpha\sim 2.5$) for $|x|<0.5\times10^{18}\,{\rm cm}$ to reflect
optically thick jet regions.

These inner regions absorb a large fraction of the emission due to the high
rest-mass density in the surrounding torus. Indeed, as the torus
rest-mass density is increased, more emission is found to be absorbed and
the optically thick region expands towards the edges of the torus
(compare the top to the bottom panels in Fig.~\ref{spix1543}). Besides
the expansion of the optically thick region, the spectral index also
generally increases for larger torus rest-mass densities. Thus, the
extent of the optically thick region and the value of the spectral index
can be used to deduce the size and the rest-mass density of the obscuring
torus.

Multifrequency VLBI observations enable us to investigate the opacity and
therefore the frequency-dependent position of the so-called ``radio
core'', \ie of the first detectable flux-density maximum. The variation
of this position with frequency provides details of the rest-mass density
and the magnetic field in the source. This effect is called
``core-shift'' \citep[see][for details]{1998A&A...330...79L}. In VLBI
observations the initial position of the jet is lost during the imaging
process and different techniques are used to reconstruct the core-shift
between adjacent frequencies \citep[see][for a
  summary]{2013A&A...557A.105F}. However, this is not needed when working
with synthetic radio maps, since the initial position is known. Here, we
define the core as the position of the first flux-density maximum along
an axial slice through the convolved radio map.

\begin{figure}[h!]
\resizebox{\hsize}{!}{\includegraphics{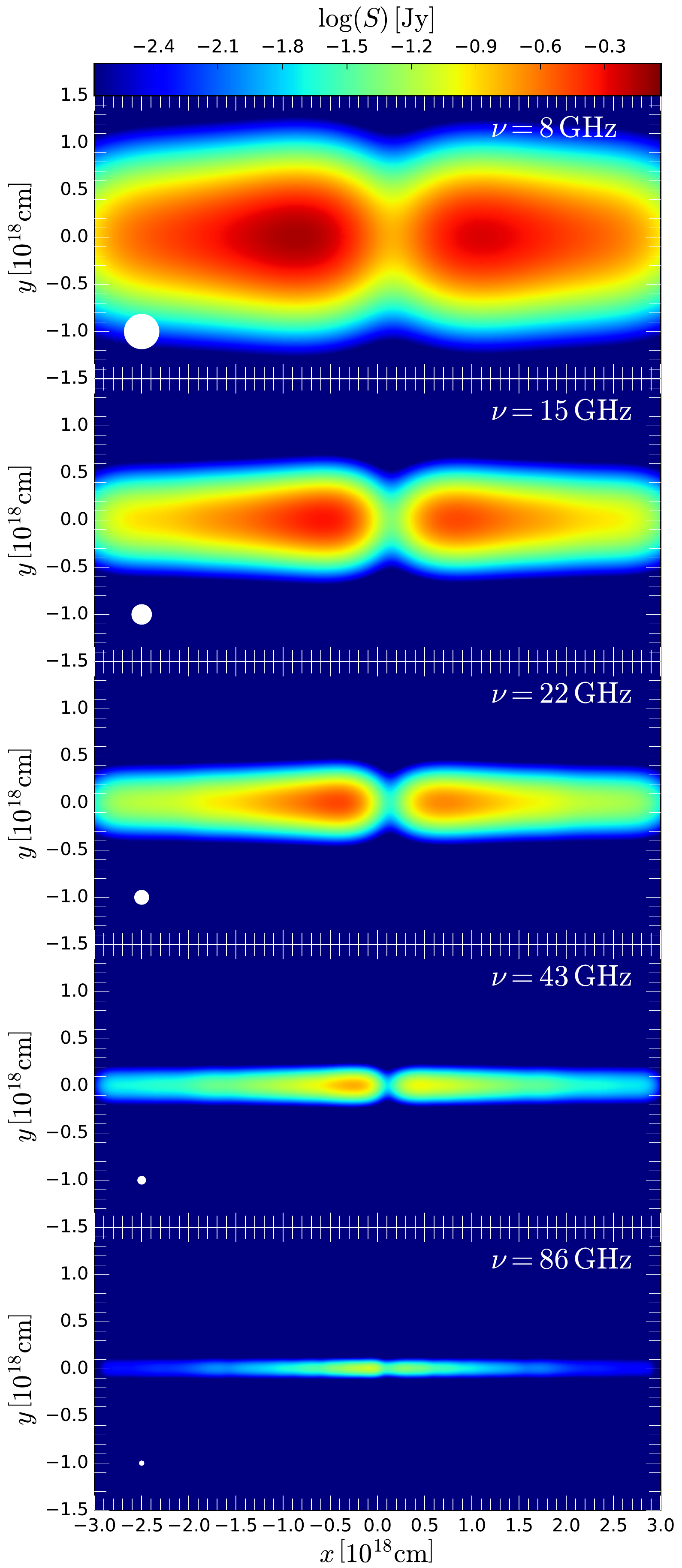}} 
\caption{Synthetic multifrequency radio maps from $8\,{\rm GHz}$ to
  $86\,{\rm GHz}$. The convolving beam is plotted in the lower left
  corner of each panel.}
\label{multifreqradio} 
\end{figure}

In Fig.~\ref{multifreqradio} we present multifrequency VLBI radio maps
from $8\,{\rm GHz}$ to $86\,{\rm GHz}$ based on our reference model (see
Table~\ref{paraem} for the parameters used). The total flux density of
the radio maps and the transversal size of the jets decreased with
frequency, where the latter could be explained by the convolution with
the observing beam (see Table \ref{beams}). Given the parameters used for
our reference model, the turnover frequency is at $\sim8\,{\rm GHz}$, so
that the flux density decreases as $S_\nu\propto \nu^{-(s-1)/2}$ for
$\nu>8\,{\rm GHz}$. Since the flux density is smaller at higher
frequencies and the sensitivity of the array also decreases with
frequency (see third column of Table~\ref{beams}), the jets appear
shorter in size at higher frequencies. Furthermore, besides a drop in the
observed emission with frequency, the gap between the jet and counter-jet
is reduced with frequency and the position of the flux-density maximum in
the jet and the counter-jet is shifted upstream.

\begin{figure}[h!]
\resizebox{\hsize}{!}{\includegraphics{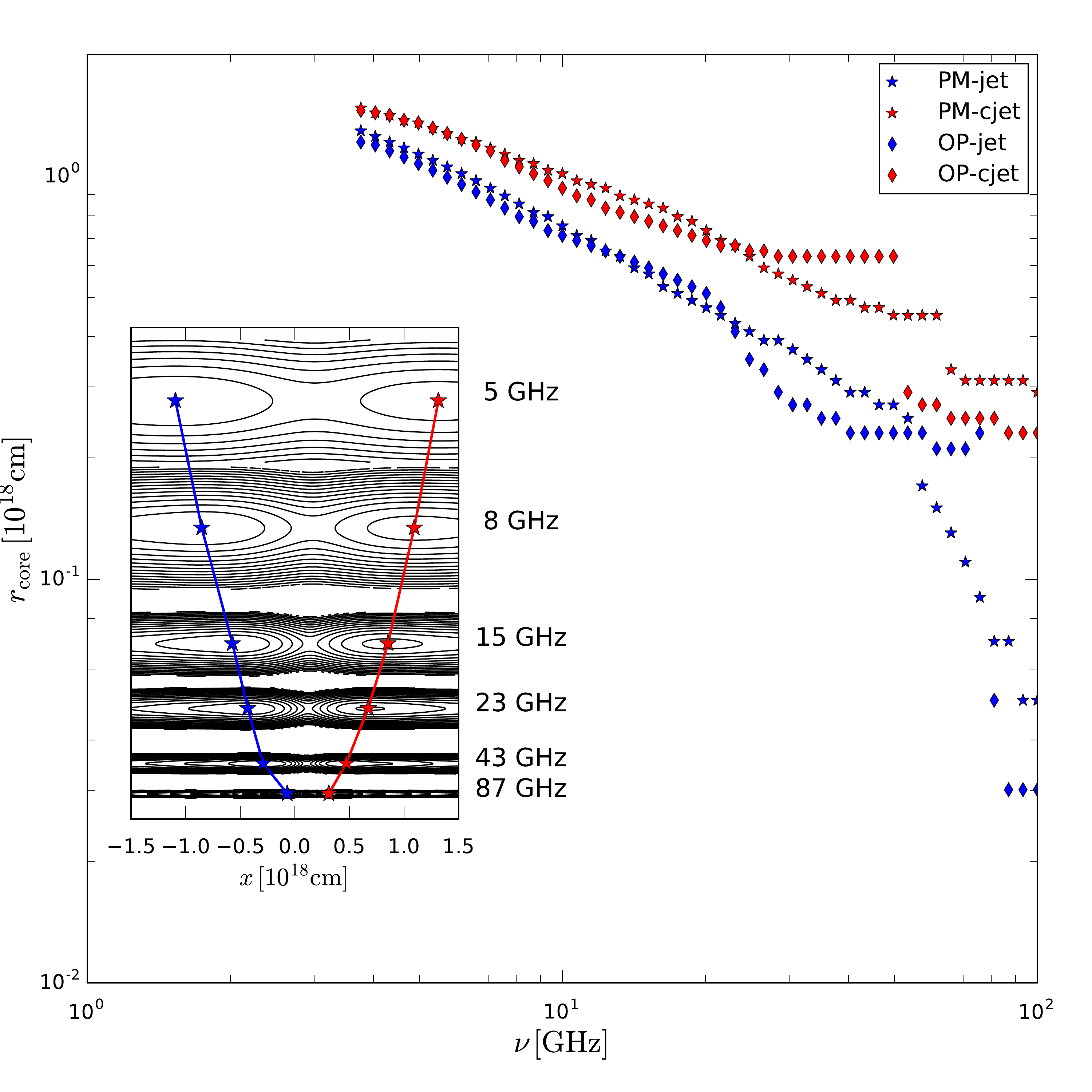}} 
\caption{Variation of the core-shift with frequency as obtained from our
  reference model. The inset panel shows a contour plot of the synthetic
  radio maps over-plotted with corresponding core position for different
  observing frequencies.}
\label{jetcoreshift} 
\end{figure}

In Fig.~\ref{jetcoreshift} we present the core-shift obtained from the
synthetic multifrequency radio maps shown in
Fig.~\ref{multifreqradio}. The red stars correspond to the position of
the radio core in the counter-jet and the blue stars to the position of
the radio core in the jet. The inset panel in Fig.~\ref{jetcoreshift}
shows a magnified view of the nozzle region of the jets and the symbols
indicate the position of the core. As mentioned previously, the core
positions in the jet and counter-jet shift upstream with frequency; at
the same time, however, the amount of the shift differs between the jet
and counter-jet. This behaviour could be explained in terms of the
different amounts of absorption along the line of sight for rays crossing
the obscuring torus. The opacity is clearly larger for rays emerging from
the counter-jet since they have to cross the entire torus, in contrast to
rays from the jet and which need to cross only a fraction of the
torus. At $\nu\sim80\,{\rm GHz}$, the torus becomes optically thin and
both cores are shifted towards the footpoint of the jets (here at
$x=0$).

Also reported in Fig.~\ref{jetcoreshift} are the core-shifts obtained
from an OP jet (diamond markers). The computed core-shifts show the same
behaviour as the PM jet for $\nu < 20\,{\rm GHz}$. For lager frequencies,
both the jet and counter-jet show a plateau occurring between $\nu \sim
20\,{\rm GHz}$ and $\nu \sim 70\,{\rm GHz}$. In this frequency range, the
recollimation shocks in the jets mimic the radio core (\ie they produce
a local flux-density maximum) and since their position is
frequency-independent, the core-shift does not decrease with
frequency. On the other hand, if the observing frequency is increased,
the upstream emission of the recollimation shocks becomes optically thin
and a new flux-density maximum is produced further upstream. Similar to
the PM jet, the torus becomes optically thin around $\nu\sim80\,{\rm
  GHz}$ and the core positions converge towards the footpoint of the jet.

\section{Summary and outlook}
\label{disc}

We investigated the connection between jets in an ambient medium of
decreasing pressure surrounded by a stationary torus. We then simulated
jets using the state-of-the art RHD code \texttt{Ratpenat}
\citep{{Perucho:2010ht}} and included a stationary torus. The properties
of the torus were motivated by both observations and theoretical
modelling. In the post-processing we used a newly developed emission code
and computed the thermal and non-thermal emission. To investigate the
effect of the scaling parameters on the observed emission (single-dish
spectra and multifrequency radio maps) we performed a detailed parameter
study and produced observable signatures which can be used to distinguish
between different jet, ambient medium and torus configurations based on
single dish and VLBI observations. We applied a core-shift analysis to
the synthetic radio maps produced and demonstrated how one can
distinguish between OP jets and PM jets depending on the slope of the
core-shift. Our simulations and the extensive analysis of the parameter
space study have shown that the observed emission is the result of a
delicate interplay between hydrodynamical effects and emission and
absorption processes.

A particularly robust result of this analysis is that the observed
spectra of PM jets exhibit higher flux densities than the spectra of OP
jets. This is due to the fact that because of the initial pressure
mismatch at the jet nozzle, the OP jets expand faster and convert
internal energy into kinetic energy more efficiently than PM jets. At
large viewing angles, the emission is strongly de-boosted for faster jets
and, as a result, less emission is received by the observer. Also, while
the spectra alone make it difficult to distinguish these two types of
jets, high-frequency radio maps show a clear difference in both the
morphology of the jet and the in the evolution of the flux density along
the jet. More specifically, OP jets exhibit local maxima in the flux
density at the position of the recollimation shocks; on the other hand,
in PM jet the flux density decreases continuously along the jet axis
\citet{1997ApJ...482L..33G, Mimica:2009de,2016A&A...588A.101F}.

Another important result of our analysis is that the inclusion of a
stationary torus in the synthetic emission consistently leads to a strong
absorption in the single-dish spectrum. This behaviour is mostly due to
the enhanced rest-mass density between the jet and the observer, which in
turn increases the opacity along the line of sight. Additionally,
depending on properties of the torus, the spectrum flattens within a
certain frequency range. Overall, the geometry of the torus has the
greatest influence on the observed spectrum. In particular, increasing
the geometrical size of the torus leads to a longer path for the rays
generated in the jets, which then pass through a high rest-mass density
field, thereby increasing the opacity along the line of sight. Indeed, if
the torus is large enough and its rest-mass density sufficiently high, it
can block a considerable fraction of the jet, which then becomes
invisible to an observer. This effect is reflected in the flattening of
the spectrum, in the appearance of an emission gap between the jet and
counter-jet in the synthetic radio maps, and in optically thick
spectral-index maps. We note that since the thermal absorption
coefficient scales like $\nu^{-3}$, the aforementioned effects are
frequency-dependent, \ie the spectra including a torus should converge to
the one without a torus for $\nu>10^{11}\,{\rm Hz}$, with the emission
gap between the jet and the counter-jet vanishing with increasing
frequency. Observationally, therefore, the spectral break could be used
to constrain the rest-mass density within the torus. Given a dense
frequency sampling, for example like that used in the FGamma program
\citep{2016arXiv160802580F}, the rest-mass density of the torus together
with additional emission parameters can in principle be fitted.

This behaviour and in particular the frequency dependence on the spectral
properties can be used to distinguish between OP and PM jets.  In
particular, since the gap is frequency-dependent, the observed onset of
the jet and counter-jet is varying with frequency. At low frequencies the
torus should block the observer's view and the onset of the jets are
separated by an emission gap; at such frequencies, OP and PM jets are
indistinguishable. However, at higher observing frequencies the gap
between the jets will decrease, exposing the differences in the two types
of jets. Since one of the key features of OP jets are recollimation
shocks, which lead to local maxima in the pressure and rest-mass density
along the jet, their presence will become visible as a local enhancement
of the emission. At a certain frequency, the onset of the jet will
coincide with the position of a recollimation shock, which itself is
frequency independent and will not change until the upstream region
becomes optically thinner and brighter than the recollimation shock. This
effect will lead to the formation of plateaus in the core-shift
plot. Because the behaviour described above is absent in the case of PM
jets, for which no strong recollimation shocks are formed and no plateaus
in the core-shift occur, observations should be able to provide
information to discriminate between these two basic classes of jets.
More specifically, using multifrequency VLBI observations, the size of
the torus can be estimated by the dimensions of the optically thick
region in the spectral index maps. The frequency-dependent size of the
emission gap can be used to obtain estimates of the rest-mass density of
the torus and analysis of the core-shift would enable one to
differentiate between OP and PM jets.

In a follow-up paper we will apply the techniques presented here to
actual observational data and include the propagation of shock waves
using the so-called ``slow-light'' approach, which will enable us to
investigate the temporal variations in the single-dish spectra and
multifrequency radio maps.

\begin{acknowledgements}

Support comes from the ERC Synergy Grant ``BlackHoleCam - Imaging the
Event Horizon of Black Holes'' (Grant 610058). MP acknowledges support by
the Spanish ``Ministerio de Econom\'{\i}a y Competitividad'' grants
AYA2013-48226-C3-2-P and AYA2015-66899-P. ZY acknowledges support from an
Alexander von Humboldt Fellowship. ER acknowledges support from the grants AYA-2012-38491-C02-01, AYA2015-63939-C2-2-P, and PROMETEOII/2014/057. CMF wants to thank Walter Alef and
Helge Rottmann for support during the computation on the MPIfR cluster
and Mariafelicia De Laurentis and Hector Olivares for
fruitful discussion and helpful comments to the manuscript.
\end{acknowledgements}

\section*{Appendix}

\begin{figure}[h!]
\resizebox{\hsize}{!}{\includegraphics{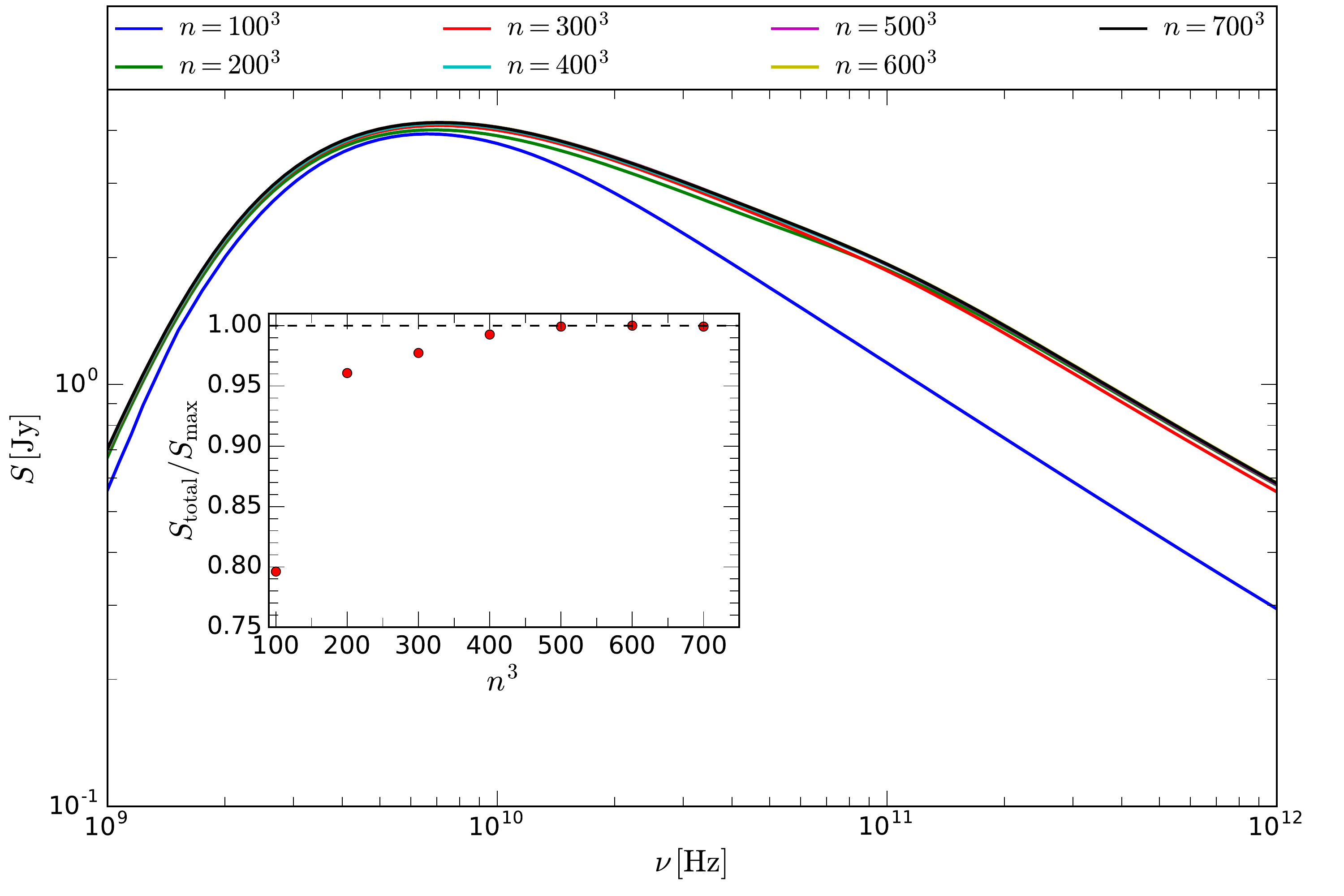}} 
\caption{Convergence test for the emission calculations. Single-dish
  spectrum for seven different resolutions.}
\label{convtest} 
\end{figure}

We performed a convergence test for our emission calculations and the
results are presented in Fig.~\ref{convtest}. For the test we used the
same initial setup and increased the number of cells used in the
computation in $100^3$ steps. In Fig.~\ref{convtest} we show the
single-dish spectrum between $1\,{\rm GHz}$ and $1\,{\rm THz}$ for
different resolutions and the inset panel shows the variation of the
total flux density with respect to the number of simulation grid
cells. The calculated flux density converges to a common value and no
significant differences was obtained above $400^3$ cells. We note that at
the lowest resolution ($100^3$) the high-frequency flux density is
significantly different compared to the high resolution runs. This
difference occurred since the high-frequency emission was created close
to the jet nozzle and therefore emanates from a region of small
transversal jet size. If the resolution was not sufficiently high to
resolve these regions, the triangulation and interpolation of these cells
onto the 3D grid failed (overestimating the contribution from the ambient
medium) and led to unphysically small flux densities.

\bibliographystyle{aa} 
\bibliography{biblo_RHD,aeireferences}

\begin{thebibliography}{29}
\expandafter\ifx\csname natexlab\endcsname\relax\def\natexlab#1{#1}\fi

\bibitem[{{Antonucci}(1993)}]{1993ARA&A..31..473A}
{Antonucci}, R. 1993, \araa, 31, 473

\bibitem[{Crusius \& Schlickeiser(1986)}]{Crusius:1986p3163}
Crusius, A. \& Schlickeiser, R. 1986, \aap, 164, L16

\bibitem[{Daly \& Marscher(1988)}]{1988ApJ...334..539D}
Daly, R.~A. \& Marscher, A.~P. 1988, \apj, 334, 539

\bibitem[{Falle(1991)}]{1991MNRAS.250..581F}
Falle, S. A. E.~G. 1991, \mnras, 250, 581

\bibitem[{{Fanaroff} \& {Riley}(1974)}]{1974MNRAS.167P..31F}
{Fanaroff}, B.~L. \& {Riley}, J.~M. 1974, \mnras, 167, 31P

\bibitem[{{Fromm} {et~al.}(2016){Fromm}, {Perucho}, {Mimica}, \&
  {Ros}}]{2016A&A...588A.101F}
{Fromm}, C.~M., {Perucho}, M., {Mimica}, P., \& {Ros}, E. 2016, \aap, 588, A101

\bibitem[{{Fromm} {et~al.}(2013b){Fromm}, {Ros}, {Perucho}, {Savolainen},
  {Mimica}, {Kadler}, {Lobanov}, \& {Zensus}}]{2013A&A...557A.105F}
{Fromm}, C.~M., {Ros}, E., {Perucho}, M., {et~al.} 2013b, \aap, 557, A105

\bibitem[{{Fuhrmann} {et~al.}(2016){Fuhrmann}, {Angelakis}, {Zensus},
  {Nestoras}, {Marchili}, {Pavlidou}, {Karamanavis}, {Ungerechts}, {Krichbaum},
  {Larsson}, {Lee}, {Max-Moerbeck}, {Myserlis}, {Pearson}, {Readhead},
  {Richards}, {Sievers}, \& {Sohn}}]{2016arXiv160802580F}
{Fuhrmann}, L., {Angelakis}, E., {Zensus}, J.~A., {et~al.} 2016, ArXiv e-prints

\bibitem[{G{\'o}mez {et~al.}(1997)G{\'o}mez, Marti, Marscher, Ibanez, \&
  Alberdi}]{1997ApJ...482L..33G}
G{\'o}mez, J.~L., Marti, J.~M., Marscher, A.~P., Ibanez, J.~M., \& Alberdi, A.
  1997, \apj, 482, L33

\bibitem[{{Hardcastle} \& {Krause}(2013)}]{2013MNRAS.430..174H}
{Hardcastle}, M.~J. \& {Krause}, M.~G.~H. 2013, \mnras, 430, 174

\bibitem[{{Hoenig}(2013)}]{2013arXiv1301.1349H}
{Hoenig}, S.~F. 2013, ArXiv e-prints

\bibitem[{Joshi \& B{\"o}ttcher(2011)}]{Joshi:2011p2764}
Joshi, M. \& B{\"o}ttcher, M. 2011, \apj, 727, 21

\bibitem[{{Kharb} {et~al.}(2008){Kharb}, {O'Dea}, {Baum}, {Daly}, {Mory},
  {Donahue}, \& {Guerra}}]{2008ApJS..174...74K}
{Kharb}, P., {O'Dea}, C.~P., {Baum}, S.~A., {et~al.} 2008, \apjs, 174, 74

\bibitem[{{Laing} \& {Bridle}(2002)}]{2002MNRAS.336..328L}
{Laing}, R.~A. \& {Bridle}, A.~H. 2002, \mnras, 336, 328

\bibitem[{Lister {et~al.}(2009)Lister, Aller, Aller, Cohen, Homan, Kadler,
  Kellermann, Kovalev, Ros, Savolainen, Zensus, \&
  Vermeulen}]{2009AJ....137.3718L}
Lister, M.~L., Aller, H.~D., Aller, M.~F., {et~al.} 2009, \aj, 137, 3718

\bibitem[{{Lobanov}(1998)}]{1998A&A...330...79L}
{Lobanov}, A.~P. 1998, \aap, 330, 79

\bibitem[{Mimica {et~al.}(2009)Mimica, Aloy, Agudo, Mart{\'\i}, G{\'o}mez, \&
  Miralles}]{Mimica:2009de}
Mimica, P., Aloy, M.~A., Agudo, I., {et~al.} 2009, \apj, 696, 1142

\bibitem[{{Mizuno} {et~al.}(2015){Mizuno}, {G{\'o}mez}, {Nishikawa}, {Meli},
  {Hardee}, \& {Rezzolla}}]{Mizuno2015}
{Mizuno}, Y., {G{\'o}mez}, J.~L., {Nishikawa}, K.-I., {et~al.} 2015, \apj, 809,
  38

\bibitem[{{Pen}(1999)}]{1999ApJS..120...49P}
{Pen}, U.-L. 1999, \apjs, 120, 49

\bibitem[{{Perucho} \& {Mart{\'{\i}}}(2007)}]{2007MNRAS.382..526P}
{Perucho}, M. \& {Mart{\'{\i}}}, J.~M. 2007, \mnras, 382, 526

\bibitem[{Perucho {et~al.}(2010)Perucho, Mart{\'\i}, Cela, Hanasz, de~la Cruz,
  \& Rubio}]{Perucho:2010ht}
Perucho, M., Mart{\'\i}, J.~M., Cela, J.~M., {et~al.} 2010, \aap, 519, A41

\bibitem[{{Porth} \& {Komissarov}(2015)}]{2015MNRAS.452.1089P}
{Porth}, O. \& {Komissarov}, S.~S. 2015, \mnras, 452, 1089

\bibitem[{{Rezzolla} \& {Zanotti}(2013)}]{Rezzolla_book:2013}
{Rezzolla}, L. \& {Zanotti}, O. 2013, {Relativistic Hydrodynamics}

\bibitem[{{Rybicki} \& {Lightman}(1986)}]{1986rpa..book.....R}
{Rybicki}, G.~B. \& {Lightman}, A.~P. 1986, {Radiative Processes in
  Astrophysics}, 400

\bibitem[{{Schartmann} {et~al.}(2005){Schartmann}, {Meisenheimer}, {Camenzind},
  {Wolf}, \& {Henning}}]{2005A&A...437..861S}
{Schartmann}, M., {Meisenheimer}, K., {Camenzind}, M., {Wolf}, S., \&
  {Henning}, T. 2005, \aap, 437, 861

\bibitem[{{Stalevski} {et~al.}(2012){Stalevski}, {Fritz}, {Baes}, {Nakos}, \&
  {Popovi{\'c}}}]{2012MNRAS.420.2756S}
{Stalevski}, M., {Fritz}, J., {Baes}, M., {Nakos}, T., \& {Popovi{\'c}}, L.~{\v
  C}. 2012, \mnras, 420, 2756

\bibitem[{{Stalevski} {et~al.}(2016){Stalevski}, {Ricci}, {Ueda}, {Lira},
  {Fritz}, \& {Baes}}]{2016MNRAS.458.2288S}
{Stalevski}, M., {Ricci}, C., {Ueda}, Y., {et~al.} 2016, \mnras, 458, 2288

\bibitem[{{Urry} \& {Padovani}(1995)}]{1995PASP..107..803U}
{Urry}, C.~M. \& {Padovani}, P. 1995, \pasp, 107, 803

\bibitem[{{van Hoof} {et~al.}(2014){van Hoof}, {Williams}, {Volk}, {Chatzikos},
  {Ferland}, {Lykins}, {Porter}, \& {Wang}}]{2014MNRAS.444..420V}
{van Hoof}, P.~A.~M., {Williams}, R.~J.~R., {Volk}, K., {et~al.} 2014, \mnras,
  444, 420

\end{thebibliography}
\end{document}